\documentclass[%
 reprint,
superscriptaddress,
 amsmath,amssymb,
 aps,
]{revtex4-2}

\usepackage{graphicx}
\usepackage{dcolumn}
\usepackage{bm}
\usepackage{xcolor}

\begin{document}


\title{Spin dynamics of a quasi-one-dimensional electron in the quantum limit}

\affiliation{Research Center for Quantum Physics, National Research and Innovation Agency, South Tangerang 15314, Indonesia}
\affiliation{Department of Physics, Tohoku University, Sendai 980-8578, Japan}
\affiliation{Department of Electrical and Electronic Engineering, Ibaraki University, Hitachi 316-8511, Japan}
\affiliation{Center for Science and Innovation in Spintronics, Tohoku University, Sendai 980-8577, Japan}
\affiliation{Takasaki Advanced Radiation Research Institute, QST, 1233 Watanuki-machi, Takasaki, 370-1292 Gunma, Japan}
\affiliation{Research Collaboration Center for Quantum Technology $2.0$, Bandung 40132, Indonesia}

\author{M.~H.~Fauzi}
\email{moha042@brin.go.id}
\affiliation{Research Center for Quantum Physics, National Research and Innovation Agency, South Tangerang 15314, Indonesia}
\affiliation{Research Collaboration Center for Quantum Technology $2.0$, Bandung 40132, Indonesia}

\author{M. Takahashi}
\affiliation{Department of Physics, Tohoku University, Sendai 980-8578, Japan}

\author{T.~Aono}
\affiliation{Department of Electrical and Electronic Engineering, Ibaraki University, Hitachi 316-8511, Japan}

\author{K. Hashimoto}
\affiliation{Department of Physics, Tohoku University, Sendai 980-8578, Japan}

\author{Y. Hirayama}
\email{yoshiro.hirayama.d6@tohoku.ac.jp}
\affiliation{Center for Science and Innovation in Spintronics, Tohoku University, Sendai 980-8577, Japan}
\affiliation{Takasaki Advanced Radiation Research Institute, QST, 1233 Watanuki-machi, Takasaki, 370-1292 Gunma, Japan}

\date{\today}

\begin{abstract}
We study electron spin dynamics whose movement is restricted to the lowest one dimensional subband channel ($G \le 2e^2/h $), through nuclear spin relaxation rate measurement ($1/T_1$). We observe an unusual double-peak structure in the $1/T_1$ profile below the lowest subband level, where the up and down spin edge channel is still largely overlap. This profile significantly deviates from the behavior predicted by a non-interacting electron model, in which the only source of relaxation is through thermal fluctuations near the Fermi level. Our experimental results, supported by theoretical calculations, suggest that enhanced electron-electron interactions at the center of a quantum point contact are the likely origin of the observed double-peak structures.

\begin{description}
\item[PACS numbers]
\end{description}
\end{abstract}

\pacs{Valid PACS appear here}
\maketitle



Quantum point contacts (QPCs) are fundamental components for the readout of an individual charge\cite{Petersson2010} and electron spin\cite{Elzerman2004} in semiconductor-based qubits. QPCs also offer a rich playground for exploring many mesoscopic physics phenomena, including interference and tunneling experiments in the quantum Hall effect\cite{DEPICCIOTTO1998395, Ji2003, Baer2014, Banerjee2018, Bartolomei2020, Nakamura2019, Nakamura2020} and many-body interactions below the lowest 1D plateau observed at zero magnetic field, colloquially known as the $0.7$ conductance anomaly \cite{Thomas, Thomas98, Sebastien2000, Kristensen2000, Reilly2002, Cronenwett, DiCarlo2006, Chung2007, Koop2007, SNakamura2009, Komijani2010, Smith2011, Iqbal2013, Bauer2013, Smith2014, Smith2015, Smith2016, MA2024}. Recent measurements using a cryogenic on-chip multiplexer architecture indicate that the barrier curvature of the QPC center is primarily determined by the random charged impurities distribution in the channel, rather than by the nominal lithographic length and width \cite{Smith2015, Smith2016, MA2024}. The distribution of the impurities modifies and complicates the barrier curvature \cite{Aono2020}. Nevertheless, nearly all measured QPCs exhibit the 0.7 anomaly with its strength varies slightly with the potential curvature. Although the exact origin of the 0.7 effect remains unclear, it is thought to involve slowly fluctuating electron spin in the QPC\cite{Schimmel2017}. These feature can emerge from enhanced electron interactions at the center of QPC \cite{Bauer2013}. Understanding the QPC properties, such as the 0.7 effect, would help us to better-readout the qubits and better engineer spin-based nanoelectronics \cite{Ensslin2006}.

However, measuring the spin dynamics related to the $0.7$ effect is notably challenging. In a typical compound semiconductor such as GaAs where such studies are carried out intensively, the electron and nuclear spin are known to couple via hyperfine interaction\cite{Hirayama_2009, Fauzi2022}. This gives us a room to study the dynamics by looking at the nuclear spin relaxation rate ($1/T_1$) since it is directly proportional to the field fluctuations strength produced by an wobbling electron spin at a nuclear site\cite{Cooper}. Although there are ample of studies on various electron spin dynamics in a compound GaAs 2D semiconductor system through nuclear spin relaxation rate $1/T_1$ measurement\cite{Tycko1460, Smet2002, Hashimoto, Kumada2006, Guo, Rhone, Guan_2015}, however to the best of our knowledge, such studies are scarcely reported in a 1D counterpart\cite{Kobayashi, Kawamura2013}. This scarcity is partly due to difficulty in generating and detecting nuclear spin polarization in a 1D, particularly in a regime where the spin degeneracy at the lowest subband is not fully lifted.

To address challenge, we recently develop a way to polarize an ensemble of nuclear spins in a quantum point contact without necessarily fully lifting the electronic Zeeman energy at the lowest 1D subband \cite{Fauzi2018}. It permits us to study electron spin dynamic in various regime where the up and down electron spin edge channel at the lowest 1D subband can be made largely overlap or fully separated both energetically and spatially by simply tuning the magnetic field strength. Connecting our study to the idea of mimicking the $0.7$ conductance anomaly observed in a quantum point contact (QPC) by applying a moderate perpendicular magnetic field is particularly appealing\cite{Shailos}, since the 0.7 anomalous line in the non-linear dc bias spectroscopy persists in a finite perpendicular magnetic field \cite{Rossler}.

\begin{figure}[t]
\begin{center}    
\centering
\includegraphics[width=\linewidth]{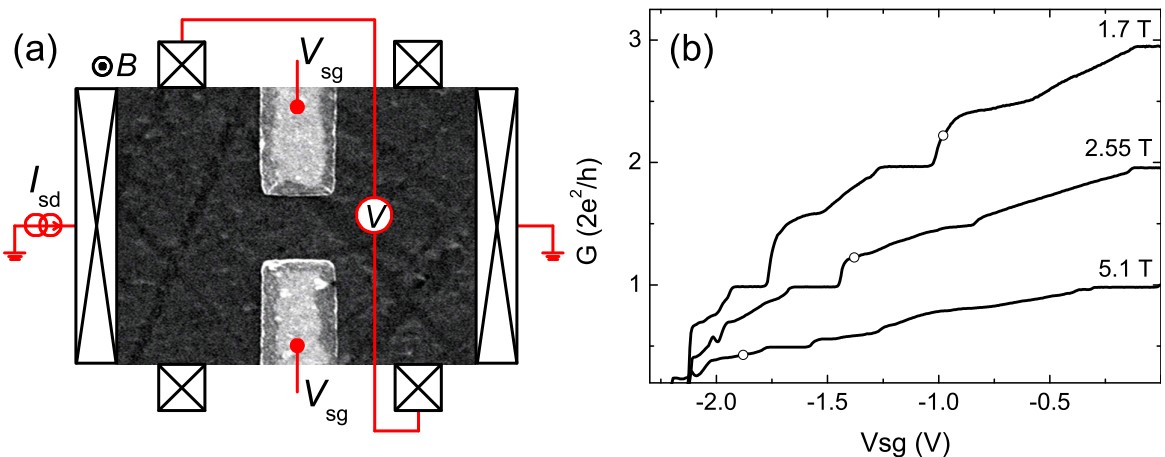}
\end{center}
\caption{(a) A gate-defined quantum point contact device and measurement setup. The lithographic gap between a pair of split metal gate is $600$ nm and the channel itself is buried deep $310$ nm below the surface. (b) conductance traces as a function of split gate bias voltage measured at $1.7$, $2.55$, and $5.1$ T perpendicular magnetic field. The fields are selected such that the bulk filling factors are an even number of $6$, $4$, and $2$, respectively. The open circle dots show current-induced dynamic nuclear polarization and detection points at each respected field.}
\label{Fig01} 
\end{figure}

Our studies are carried out on a $20$-nm wide GaAs quantum well with the center of our two dimensional electron gas (2DEG) is located $310$ nm below the surface. The electron density at low temperature is controlled by a back gate ($V_{\rm{BG}}$). The layout of the present device and measurement setup are displayed in Fig. \ref{Fig01}(a). Applying a negative bias to the two Schottky gates ($V_{\rm{SG}}$), with a lithographic gap of $600$ nm and $500$ nm long, depletes the electrons underneath and defines a 1D constriction. We measure the diagonal voltage to probe the number of transmitted modes (or spin edge channels at a finite magnetic field) through the point contact. Throughout the measurement process, we always equally bias the pair of split gates.

Fig. \ref{Fig01}(b) displays the conductance traces as a function of $V_{\rm{SG}}$ measured at a field of $1.7$, $2.55$, and $5.1$ T. The fields are chosen in such a way that the bulk carriers a pair of spin edge modes (up and down electron spin) before entering the constriction\cite{Fauzi2018}. We set the split gate bias to a certain value shown in the open dots in Fig. \ref{Fig01}(b), where the inner most spin down edge channel is completely reflected, to be able to dynamically polarize the nuclear spins in the point contact.

\begin{figure}[t]
\begin{center}    
\centering
\includegraphics[width=\linewidth]{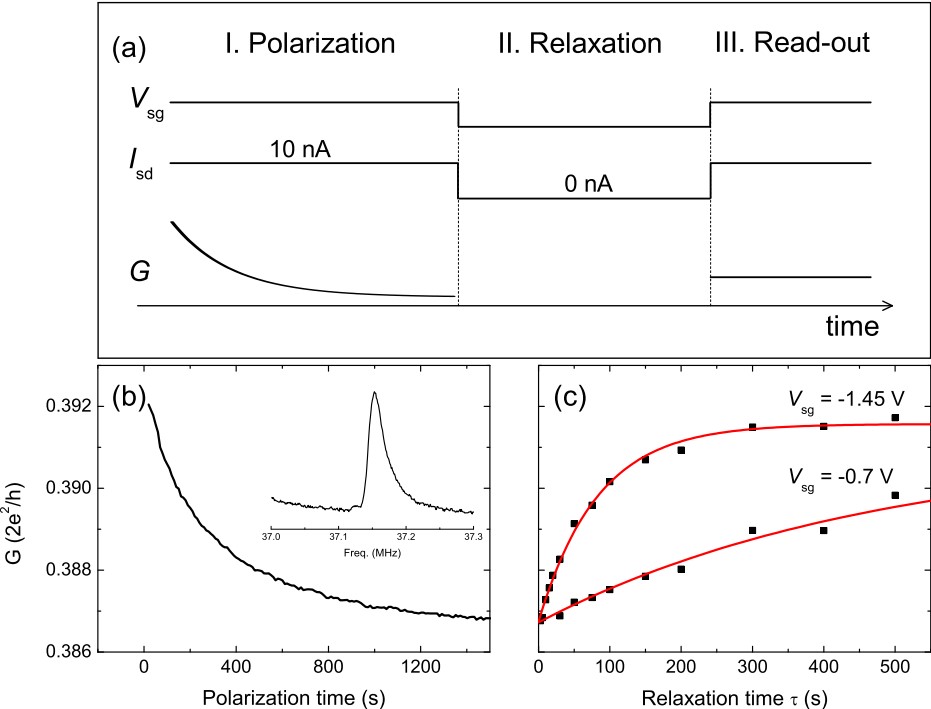}
\end{center}
\caption{(a) Pump-probe timing sequence to measure nuclear spin relaxation rate, $1/T_1$. (b) Generating dynamic nuclear spin polarization in the point contact. Inset shows the $^{75}$As RDNMR signal measured at a field of $5.1$ T, proving that the conductance change is nuclear spin related. (c) Example of nuclear spin relaxation curves due to interaction with two different electronic states in the quantum point contact ($V_{\rm{sg}} = -0.7$ and $V_{\rm{sg}} = -1.45$ V) measured at a field of $5.1$ T. The red line is a single exponential fit to the experimental data (square dots) from which $1/T_1$ is extracted.}
\label{Diagram} 
\end{figure}

Fig. \ref{Diagram}(a) displays the pump-probe timing sequence to measure nuclear spin relaxation rate $1/T_1$, adopting initial protocol developed in Ref. \cite{Hashimoto, Kumada2006}. First, we initialize/polarize the nuclear spins in the point contact by applying $I = 10$ nA to induce dynamic nuclear polarization (DNP). We flow the current and wait for as long as $1500$ seconds to reach a steady state where the conductance is already saturated as displayed in Fig. \ref{Diagram}(b). Note that we always carefully check whether the conductance change is nuclear spin in origin or not through NMR measurement as shown in the inset of Fig. \ref{Diagram}(b) by applying rf magnetic field across the Larmor frequency $^{75}$As nuclei. Second, we tune the point contact to a desired electronic state in the lowest 1D subband by changing the gate voltage value ($V_{\rm{SG}}$) for a certain time interval $\tau$. During the second step, we turn the source-drain current off to avoid current-induced DNP and to ensure the electron is in an equilibrium state. Finally, we revert the state to the initial step condition to read out the remaining nuclear spin polarization by monitoring the conductance change. To get a full nuclear spin relaxation curve, we repeat the last two steps for different time interval $\tau$. Fig. \ref{Diagram}(c) displays an example of nuclear spin relaxation curves due to interaction with two different electronic states, controlled by $V_{\rm{SG}}$, from which $1/T_1$ of about $2.1 \times 10^{-3}$($1.2 \times 10^{-2}$) s$^{-1}$ for $V_{\rm{SG}} = -0.7$($V_{\rm{SG}} = -1.45$) V is extracted.

\begin{figure}[t]
\begin{center}    
\centering
\includegraphics[width=\linewidth]{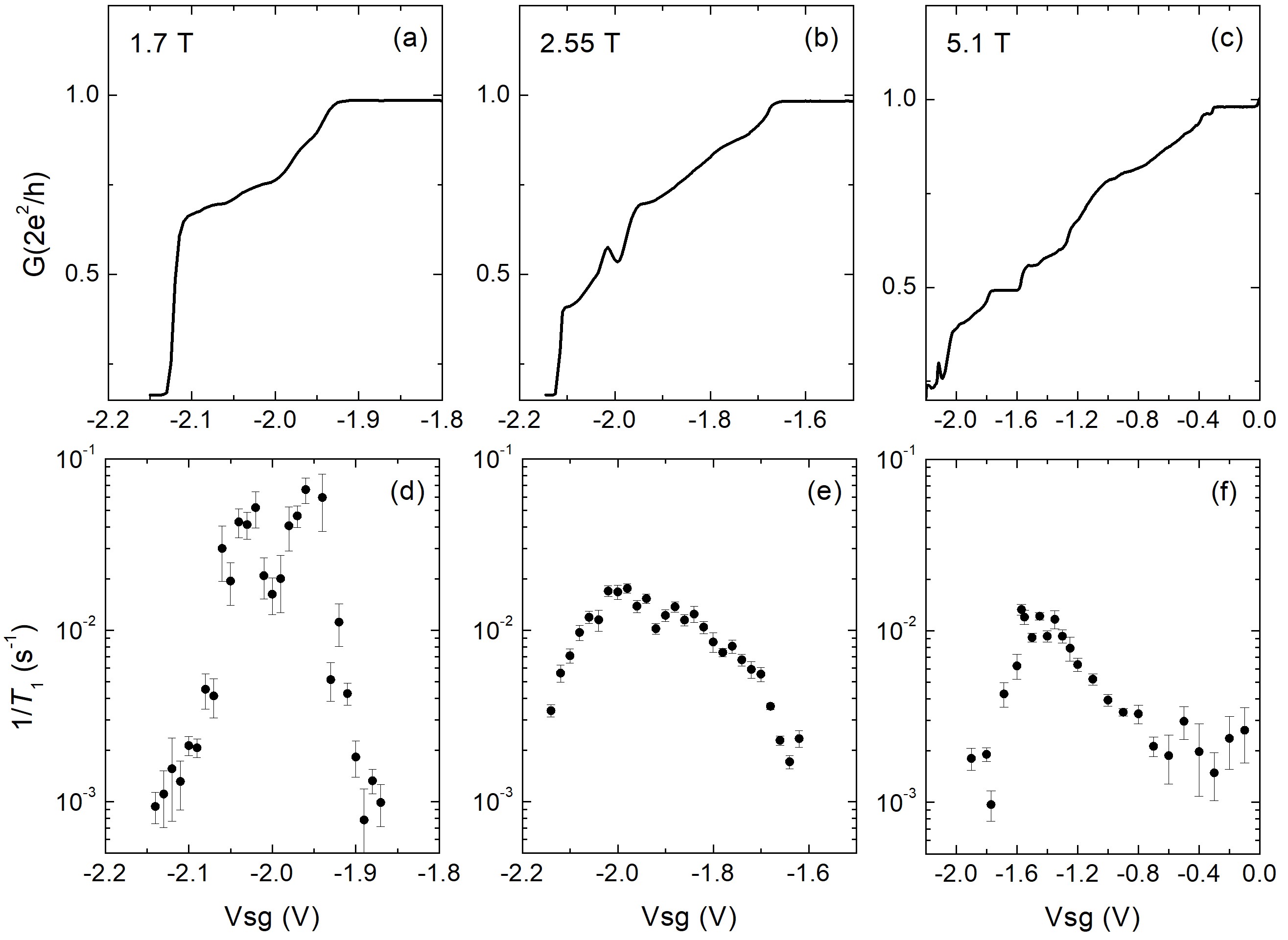}
\end{center}
\caption{(a)-(c) The lowest subband conductance traces as a function of split gate voltage ($V_{\rm{sg}}$) measured at three different perpendicular magnetic fields. (d)-(f)  The corresponding semi-log plot of nuclear spin relaxation rate ($1/T_1$) profiles. All presented data are taken at a temperature of 100 mK and at electron density $n = 2.83 \times 10^{15}$ m$^{-2}$.}
\label{Fig02} 
\end{figure}

Fig. \ref{Fig02}(a)-(c) displays the conductance traces as a function of split gate bias voltage ($V_{\rm{sg}}$) at the lowest 1D subband ($G \le 2e^2/h$) selectively measured at three different magnetic fields. At a field of $1.7$ T, one can see a structure below $G < 2e^2/h$ since the up and down electron spin edge channel begins to split but still largely overlap both energetically and spatially. This is believed to mimic the $0.7$ feature with interaction-induced spin gap\cite{Shailos}. Stepping up the field to $2.55$ T, the overlap gets smaller and the conductance structure gets more complicated with the appearance of a bump around $V_{\rm{sg}} = -2$ V. Finally, the two spin edge channels are fully separated at a field of $5.1$ T with additional development of several fractional edge channels\cite{Rossler}.

Let us start off by first discussing the nuclear spin relaxation rate profile for the highest measured field of $5.1$ T plotted in a semi-log scale, as shown in Fig. \ref{Fig02}(f). Initially when the first subband is still fully occupied, corresponding to $G = 2e^{2}/h$,  $1/T_1$ value is relatively small on the order of $10^{-3}$ s$^{-1}$. It is reasonable because electron spin excitations are highly suppressed by a large cyclotron gap, about three order of magnitude larger than the thermal energy. In this case, the only available nuclear spin relaxation channel is through diffusion. With applying more negative bias to the spit gates, the relaxation rate starts increasing linearly (or exponentially in a linear scale) up to $V_{\rm{SG}} = -1.35$ V, reaching a value of $1.16 \times 10^{-2}$ s$^{-1}$. This is followed by a flat response for a narrow $V_{\rm{SG}}$ region before it gets inflected and its value drops down to $4.28 \times 10^{-3}$ s$^{-1}$ due to the presence of the Zeeman gap upon reaching the half-integer plateau, centered at $V_{\rm{SG}} = -1.685$ V. The value continues to drop since we have less and less electrons to participate in the relaxation process toward the pinch off point. The relaxation rate value in the absence of electrons reported here is consistent with the earlier report measured in a 2D\cite{Hashimoto} and a 1D GaAs system\cite{Kawamura2013}. 

Furthermore, we do not see a dramatic enhancement of nuclear spin relaxation rate in the vicinity of the half-integer plateau as reported in a $3$-$\mu$m long quantum wire by Kobayashi et al\cite{Kobayashi}, leading us to rule out the formation of low-frequency collective electron spin excitations such as Skyrmion. Considering that the effective channel length of our 1D constriction at the lowest subband is about $36$ nm (see the supplementary material on how to determine the length), much smaller than the expected Skyrmion diameter of about $176$ nm for an ideal 2DEG with zero thickness\cite{Ezawa}. We conclude it is unlikely for such collective excitation to stably form. The electron spin polarization itself at local filling factor 1 can be reduced by $30\%$ from unity due to electron-nuclear spin flip scattering \cite{Chida2012}, contributing to an increase in the nuclear spin relaxation rate. 

Let us know move on to discuss the relaxation rate profile obtained at the field of $2.55$ T as displayed in Fig. \ref{Fig02}(e). $1/T_1$ profile increases linearly with increasing a negative bias voltage to the $V_{\rm{SG}}$ before reaching a maximum value of $1.75 \times 10^{-2}$ s$^{-1}$ at $V_{\rm{sg}} = -1.98$ V and a little dip at $V_{\rm{SG}} = -1.92$ V. Compare to the data at $5.1$ T, we do not see a significant increase in the peak height as it should be when the magnetic field is lowered the field to $2.55$ T. However when we lower the field to $1.7$ T, the peak height of $1/T_1$ increases by nearly an order of magnitude, as depicted in Fig. \ref{Fig02}(d). While an increase in the peak height of $1/T_1$ is anticipated, the dip in the relaxation rate profile on the other hand becomes more pronounced. $1/T_1$ increases up to $6.6 \times 10^{-2}$ s$^{-1}$ at $V_{\rm{SG}} = -1.96$ V and decreases down to $1.6 \times 10^{-2}$ s$^{-1}$ at $V_{\rm{SG}} = -2.0$ V with $G \approx 0.76\times2e^2/h$ before it recovers up to $5.2 \times 10^{-2}$ s$^{-1}$ at $V_{\rm{SG}} = -2.02$ V, creating a double peak structure. This phenomenon is clearly unexpected for a non-interacting electron \cite{Cooper}. A similar double-peak structure in the $1/T_1$ profile has been observed by Kawamura et al. in a single quantum dot \cite{Kawamura2013}. They attributed their finding to a Kondo-like resonance state formed in the quantum dot. Moreover, the asymmetric noise factor observed in shot noise measurement around the 0.7 conductance stems from the spin-polarized electrons in the QPC \cite{DiCarlo2006}.

\begin{figure}[ht]
\begin{center}    
\centering
\includegraphics[width=\linewidth]{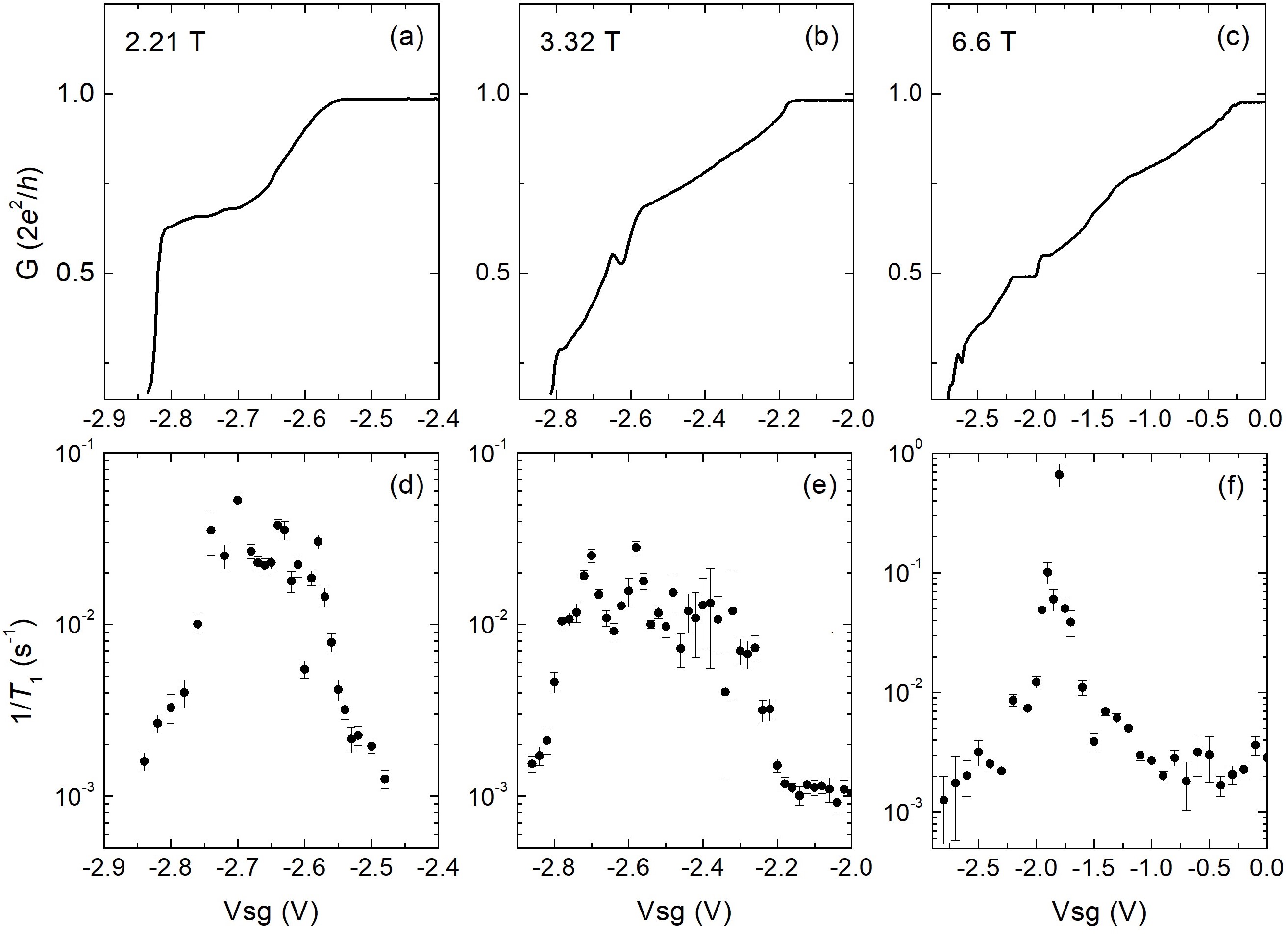}
\end{center}
\caption{(a)-(c) The lowest subband conductance traces as a function of split gate voltage ($V_{\rm{sg}}$) measured at three different perpendicular magnetic fields. (d)-(f)  The corresponding semi-log plot of nuclear spin relaxation rate ($1/T_1$) profiles. All presented data are taken at a temperature of 100 mK and at electron density $n = 3.52 \times 10^{15}$ m$^{-2}$.}
\label{T1_BG6V} 
\end{figure}

Displayed in Fig. \ref{T1_BG6V}, we measure similar $1/T_1$ profile but at higher 2DEG electron density $n = 3.52 \times 10^{15}$ m$^{-2}$ by tuning the back gate voltage to $V_{\rm{BG}} = 6$ V. The same conductance plateau now appears at higher magnetic field than the ones shown in Fig. \ref{Fig02}, as expected. The double peak structure in the relaxation profile observed in Fig. \ref{Fig02} (d) appears also in Fig. \ref{T1_BG6V} (d), except that the dip ($1/T_1 = 2.2 \times 10^{=2}$ s$^{-1}$ at $V_{\rm{SG}} = -2.66$ V with $G \approx 0.73\times2e^2/h$) which defines the double structure is noticeably less pronounced. For an intermediate case displayed in Fig. \ref{T1_BG6V}(e) and Fig. \ref{Fig02}(e), the profile around $G \approx 0.7\times2e^2/h$ is quite similar.

We argue that the appearance of the double peak structure in the nuclear spin relaxation rate $1/T_1$ in Fig. \ref{Fig02}(d) and \ref{T1_BG6V}(d) arises from variations in the effective Zeeman energy, which combines a bare Zeeman energy $Z_e$ and an electron-electron interaction terms. The $0.7$ anomaly is accompanied by an enhanced $g$-factor in finite magnetic fields due to electron-electron interactions \cite{Thomas98}. Bauer et al., in Ref. \cite{Bauer2013}, thoroughly explore the $0.7$ structure issue using the Hubbard model, including the mechanism behind the $g$-factor enhancement caused by the Hubbard interaction. They also show that the gate voltage dependence of the spin susceptibility $\chi_{\rm tot}$ peaks near $G \sim 0.7 \times 2e^2/h$, with the peak height increasing as the Hubbard interaction strengthens. In the mean-field approximation of the Hubbard model, taking into account the Zeeman field, the gate voltage dependence of the magnetization $m$ near the center of the QPC exhibits a similar peak structure \cite{Kawamura2015}.  In this case, the effective Zeeman energy (spin gap) near the QPC center is 
\begin{equation}  \label{eq:effective-zeeman}
 \tilde{Z_{\rm e}} = Z_e + U m
\end{equation}
with the strength of the Hubbard interaction $U$. At high magnetic fields, the bare Zeeman energy dominates, while at lower magnetic fields, the interaction term becomes more prominent, resulting in a suppressed spin flip rate at the peak and increased spin flip rate on either side, creating the observed double-peak profile (see the supplementary material for detail calculation).

Now moving on to the $1/T_1$ profile for the fully open Zeeman gap case shown in Fig. \ref{Fig02}(f), we see a dramatic enhancement of $1/T_1$ value reaching $0.66$ s$^{-1}$ at $V_{\rm{SG}} = -1.8$ V at the right flank of the half-conductance plateau. We attribute this enhancement due to a Skyrmion formation since now the diameter is sufficiently small to fit inside the QPC as we increase the magnetic field to $6.6$ T. We observe no enhancement in the left flank as in Ref. \cite{Hashimoto, Kobayashi} presumably since the channel is already too narrow for the Skyrmion to stably form.

In summary, we have measured the nuclear spin relaxation rate in the lowest subband level of a quantum point contact, both when the spin gap is fully open and when it is partially open. We find that the relaxation rate is consistent with a non-interacting case when the spin gap is fully open, where nuclear spin relaxation occurs through thermal spin fluctuations near the Fermi energy. However, when the spin gap is partially open, the relaxation rate deviates substantially from the non-interacting picture. Our model calculation based on a finite-magnetization in the QPC can explain the observed double peak structure in the $1/T_1$ profile. The finite magnetization stems from the enhanced electron-electron interactions at the center of QPC. Our $1/T_1$ measurements reveal that the spin dynamics can differ considerably even with a seemingly similar conductance profile. This makes it useful to identify electronic phenomena where conductance measurements alone fail to provide adequate information.

We would like to thank K. Muraki of NTT Basic Research Laboratories for supplying high quality wafers for this study. M.H.F, M.T. and Y.H. acknowledge support from the Graduate Program in Spintronics, Tohoku University. T.A. acknowledges financial support from KAKENHI Grants Nos. JP20K03814 and JP24K06912. K. H. acknowledges financial support from KAKENHI Grants No. 23K23165. Y.H. acknowledges financial support from KAKENHI Grants No. 18H01811.

\pagebreak
\widetext
\begin{center}
\textbf{\large Supplementary for Spin dynamics of a quasi-one-dimensional electron in the quantum limit}
\end{center}

\setcounter{equation}{0}
\setcounter{figure}{0}
\setcounter{table}{0}
\setcounter{page}{1}
\makeatletter
\renewcommand{\theequation}{S\arabic{equation}}
\renewcommand{\thefigure}{S\arabic{figure}}
\renewcommand{\bibnumfmt}[1]{[S#1]}
\renewcommand{\citenumfont}[1]{S#1}

\section{Bias cooling}

To suppress random telegraph noise \cite{ARLong}, we performed a bias-cooling technique during the cooldowns from room to a base temperature by continuously applying $+0.4$ V to the pair of split metal gates while Ohmic contac pads and the back/bottom gate were all grounded. To avoid accidentally applying an excessive current during the bias-cooling, we set the current compliance to $100$ nA.


\begin{figure*}[b]
\begin{center}    
\centering
\includegraphics[width=0.75
\linewidth]{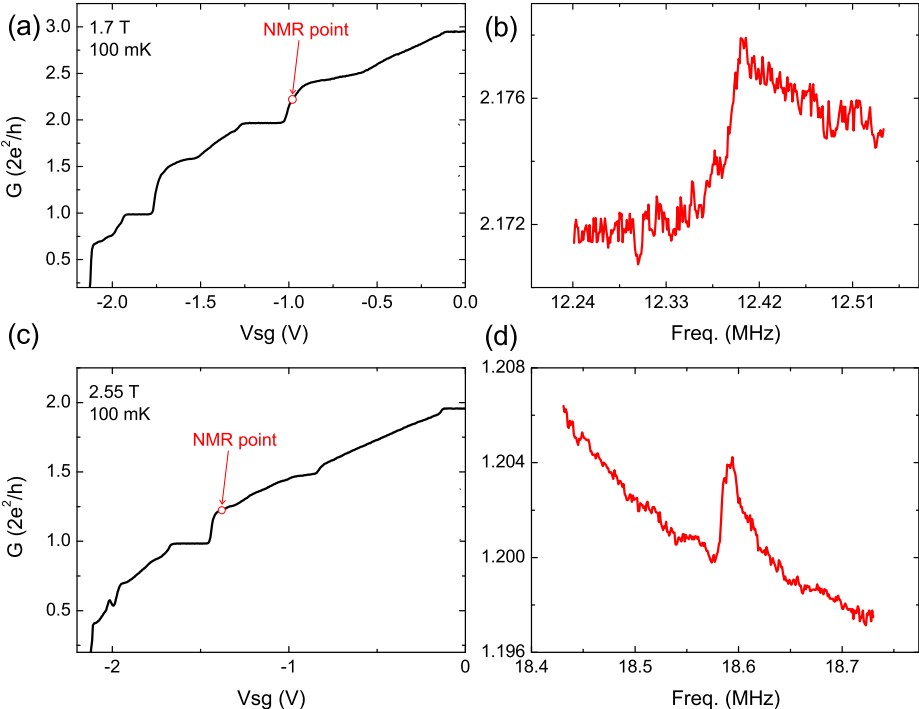}
\end{center}
\caption{(a) magneto-transport traces as a function of split gate bias measured at a field of $1.7$ T and $100$ mK temperature. The NMR operating point is indicated by a red dot at which the NMR spectra displayed in panel (b) was measured. (c) similar traces but taken at a field of $2.55$ T. The corresponding NMR traces is displayed in panel (d).}
\label{FigS01} 
\end{figure*}

\section{RDNMR signal}

Before undertaking nuclear spin relaxation rate measurement displayed in the main text at each respective magnetic field, we consistently verified that the resistance changes used to evaluate nuclear spin relaxation rate in the main text were nuclear spin related. We confirmed that this was the case since $^{75}$As RDNMR signals were visibly detected at each selected operating points as evidenced in Fig. 1.

\section{DC bias spectroscopy}

\begin{figure*}[t]
\begin{center}    
\centering
\includegraphics[width=0.75
\linewidth]{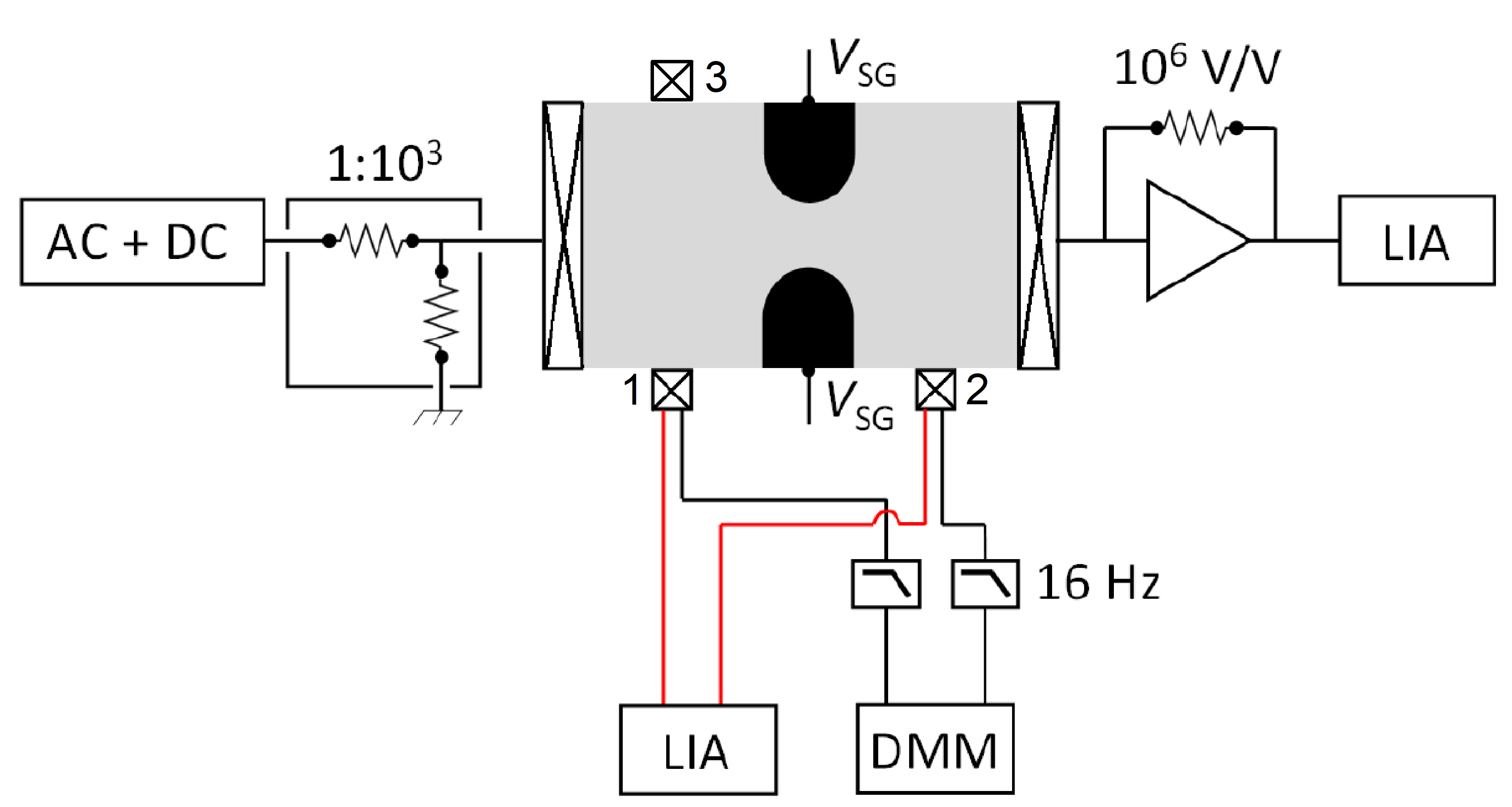}
\end{center}
\caption{Schematic of non-linear dc spectroscopy measurement setup. AC and DC voltage drop across the point contact are simultaneously measured using a lock-in amplifier (LIA) and a digital multimeter (DMM). A current flow to the drain contact is pre-amplified by a trans-impedance with a gain set to $10^{6}$ V/V. When the perpendicular magnetic field is applied, Ohmic contact pad 1 and 3 are used instead to measure the voltage drop diagonally.}
\label{FigS02} 
\end{figure*}

\begin{figure}[t]
\begin{center}    
\centering
\includegraphics[width=
\linewidth]{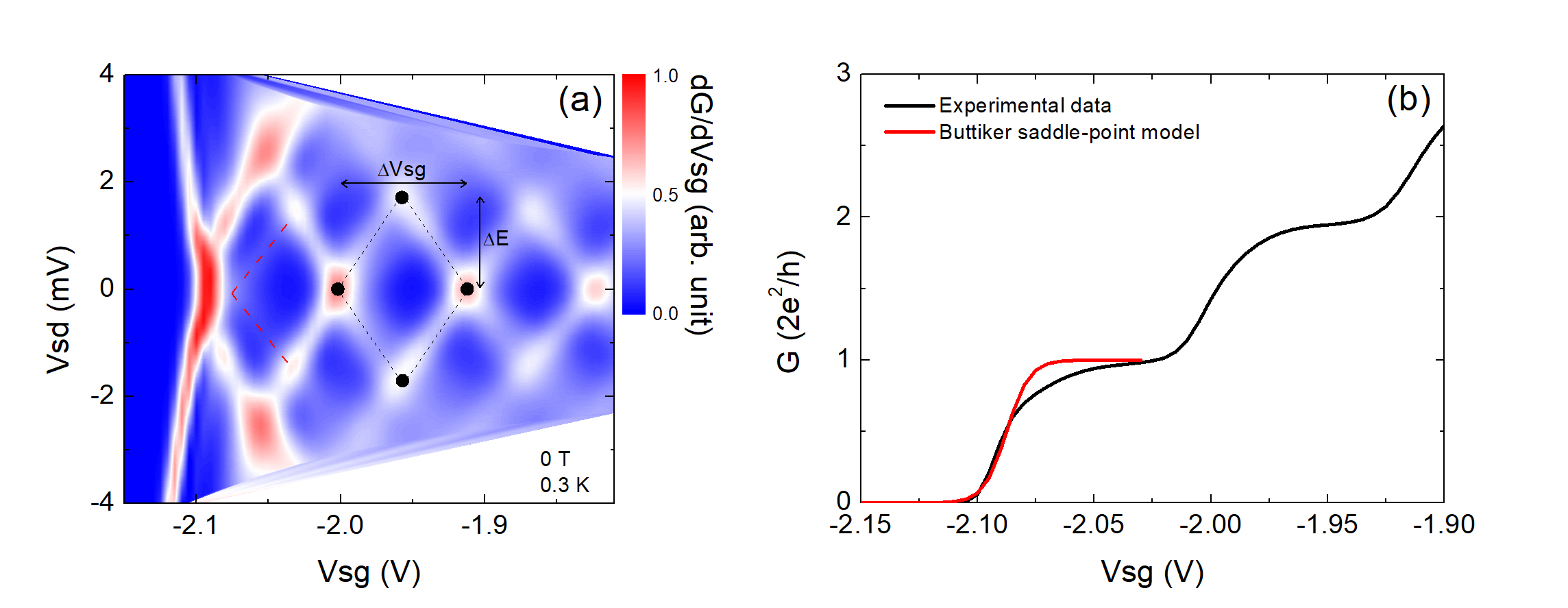}
\end{center}
\caption{(a) Color-coded magneto trans-conductance traces measured at zero field in a 300 mK single shot fridge. The lowest subband spacing is estimated to be about $2.813 \pm 0.1$ meV. The red dashed line highlights anomalous 0.7 conductance. (b) Measured differential conductance (black) can be fitted by a saddle-point model (red). The best fit to the measured curve for the lowest conductance plateau gives $\hbar \omega_{x} = 0.85$ meV ($l_{x} \approx 36$ nm).}
\label{FigS03} 
\end{figure}

To do dc bias spectroscopy measurement schematically displayed in Fig. \ref{FigS02}, we supplied AC and DC voltage signal to the device from a NF signal generator WF 1947 with $1:10^{3}$ voltage divider ratio. AC voltage fed to the device was about $80$ $\mu$V ($< 4KT$ at $T = 300$ mK) with a frequency set to $77$ Hz, which then was used to frequency-lock two lock-in amplifiers. We pre-amplified the ac current signal before feeding it to a Signal Recovery 7265 digital lock-in amplifier using a Femto DLCPA-200 current preamplifier with a gain set to $10^{6}$ V/V. AC and DC voltage drops across the point contact were simultaneously recorded by the 7265 lock-in amplifier and a DMM Keithley 34411A, respectively. We inserted a single-pole RC line filter with a cut-off frequency of about $16$ Hz ($R = 100$ k$\Omega$, $C = 0.1$ $\mu$F) in between the lock-in and the DMM to cut a cross-talk. Note that the spectroscopy measurements were carried out in a separate cooldown in a 300 mK single shot fridge.  The pinch-off voltage was almost the same as the ones shown in the main text .

Fig. \ref{FigS03}(a) displays non-linear dc bias spectroscopy measurement to extract several parameters including subband energy spacing and lever arm, in particular for the lowest subband. The first and second subband is separated by about $\Delta E \approx 2.813 \pm 0.1$ meV. This corresponds to a change in gate voltage of about $\Delta V_{\rm{sg}} \approx 0.096$ V. The lever arm relating the energy spacing to the change in gate voltage is $\Delta E/\Delta V_{\rm{sg}} \approx 29.3$ meV/V. The extracted parameter is then used as an input parameter to fit the lowest differential conductance plateau with a saddle-point model as displayed in Fig. \ref{FigS03}(b), giving QPC longitudinal barrier curvature potential $\hbar \omega_{x} = 0.85$ meV (longitudinal barrier curvature length $l_{x} \approx 36$ nm). The lowest plateau is deviated from the sadlle-point model and is believed due to the $0.7$ effect. The deviation from a non-interacting model is observed also in the non-linear dc bias spectroscopy displayed in Fig. \ref{FigS03}(a) as a line intersecting the first diamond highlighted by the red dashed line.

\begin{figure*}[t]
\begin{center}    
\centering
\includegraphics[width=0.75
\linewidth]{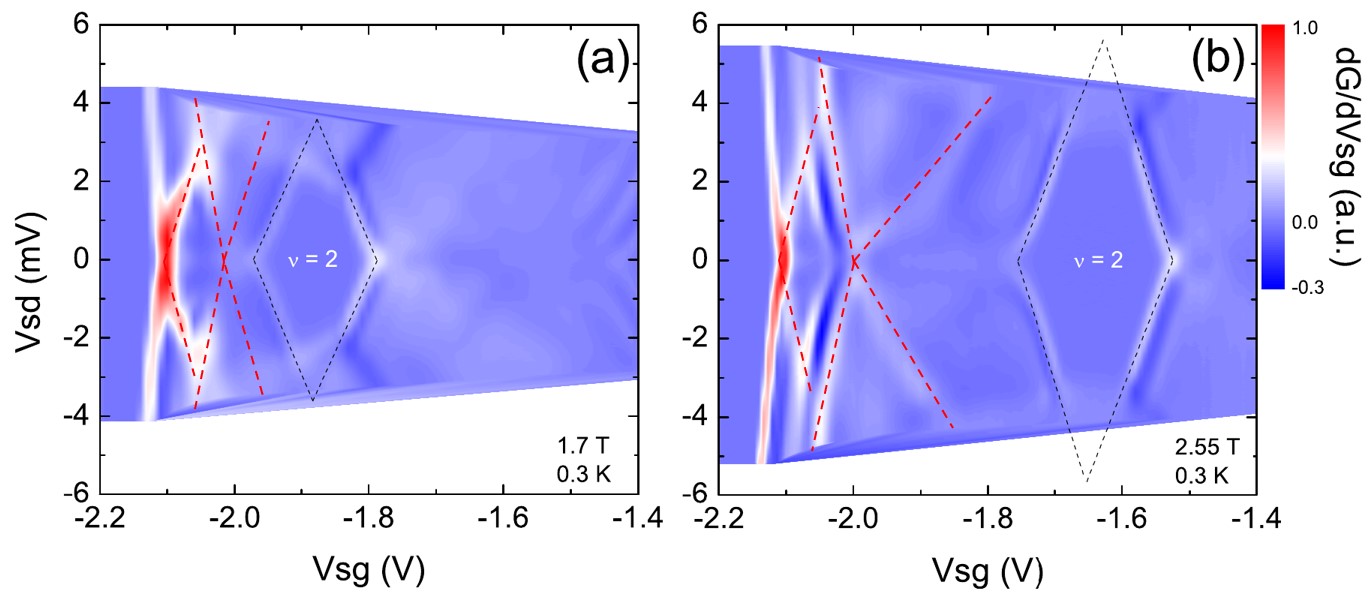}
\end{center}
\caption{Color-coded magneto trans-conductance traces measured at (a) $1.7$ T and (b) $2.55$ T field in a 300 mK single shot fridge. The lowest magneto-electric subband spacing ($\nu = 2$) for $1.7$ ($2.55$) T is about $4.047 \pm 0.17$ ($5.6624 \pm 0.4$) meV.  The red dahsed line corresponds to the anomaluous line which shares the same origin with the 0.7 anomalous line at zero magnetic field.}
\label{FigS04} 
\end{figure*}

Fig. \ref{FigS04} shows similar DC bias spectroscopy, but taken at a finite magnetic field applied perpendicular to the 2DEG plane. The magneto-electric subband spacing and the electronic Zeeman gap at each magnetic field can be directly extracted from the data. The diamond shape marked by the black dashed line does not appear to be fully developed and is truncated when one of the chemical potentials (source or drain) approaches the subband level, possibly due to an increase in inter-edge scattering events. Furthermore, we observe an anomalous line highlighted by the red dashed line, which we believe shares the same origin as the 0.7 anomalous line at zero magnetic field. The persistence of this anomalous line at $1.7$ and $2.55$ T suggests the presence of 1D transport close to that observed at zero magnetic field.

\begin{figure*}[t]
\begin{center}    
\centering
\includegraphics[width=0.5
\linewidth]{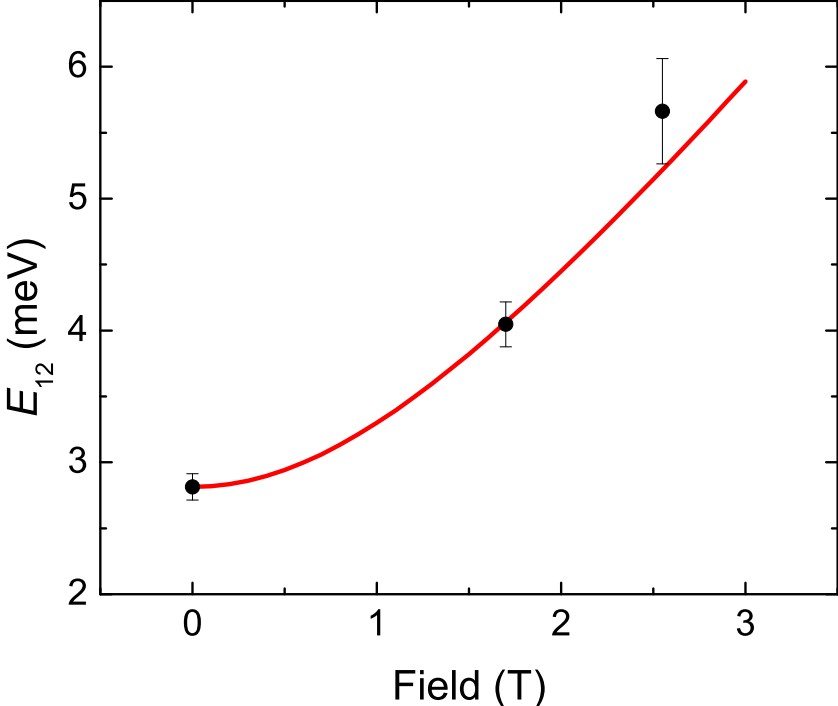}
\end{center}
\caption{The lowest magneto-electric subband plotted as a function of magnetic field extracted from the data presented in Fig. 3 and 4. The red line is the expected magneto-electric energy.}
\label{Fig04} 
\end{figure*}

We plot the lowest subband spacing in Fig. \ref{Fig04} and compare them to the non-interacting energy spectrum of a quantum point contact in the presence of perpendicular magnetic field $B$ (red solid line)
\begin{equation} \label{eq:1}
E_{12} = \hbar\sqrt{\omega_y^{2} + \omega_c^{2}}
\end{equation}
here $\omega_y$ and $\omega_c = eB/m^{*}$ are lateral electrostatic confinement and cyclotron frequency, respectively.

\section{Nuclear spin relaxation rate calculation}

To understand the origin of the double peak structure in the $1/T_1$ profile, we compute the nuclear spin relaxation rate using the following equation [33]
\begin{equation}  \label{eq:cooper-formula}
 T_1^{-1} = \Gamma_0 \int_{|Z_e|/2}^{\infty } d\epsilon\, \frac{  f(\epsilon)[1-f(\epsilon)]  }{\sqrt{\epsilon^2 - (Z_e/2)^2} }
\end{equation}

where $\Gamma_0$ is a characteristic rate that is determined by the hyperfine coupling constant and the geometry of the QPC. Typical GaAs QPC, $\Gamma_0 \simeq 0.5$ Hz. 
$Z_{\rm e} = g \mu_{\rm B} B$ is the Zeeman energy
(note that $g = - |g|$ with $|g|=0.44$ for the bulk GaAs.)

\begin{figure}[ht]
\begin{center}    
\centering
\includegraphics[width=0.5\linewidth]{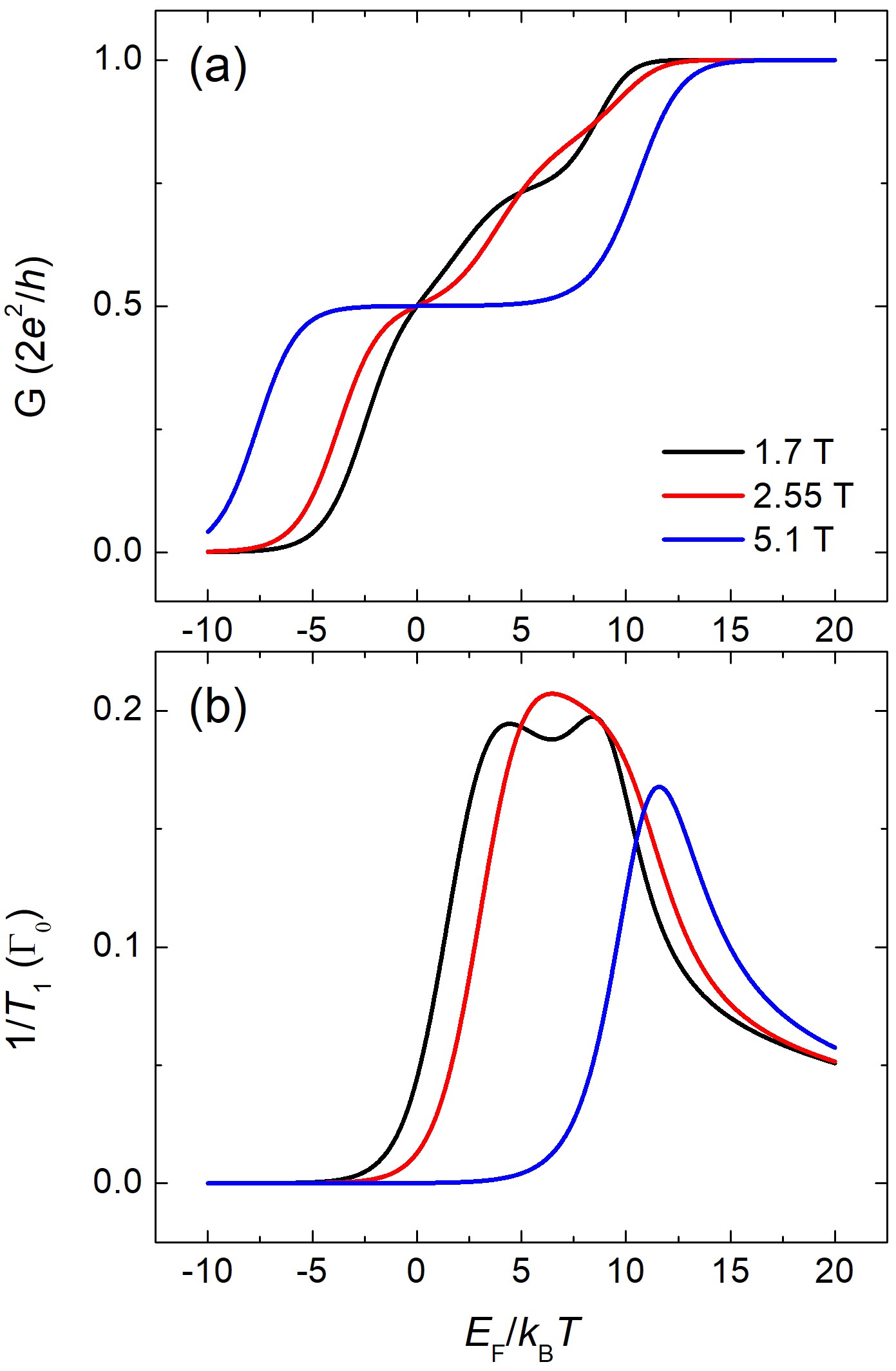}
\end{center}
\caption{(a) Calculated conductance as a function of $E_{\rm{F}}/k_{\rm{B}}T$ for $B = 1.7,~2.55~\rm{and}~ 5.1$ T. (b) Calculated nuclear spin relaxation rate for each respective magnetic field. The temperature $T$ is $100$ mK with coefficients $u/k_{\rm{B}}T = 5,~ 4.5,~3$, $\sigma/k_{\rm{B}}T=5,~5.5,~11$ and $E_0/k_{\rm{B}}T=9,~11,~12$. See the text for details.}
\label{Model} 
\end{figure}
 
The calculation model for three different magnetic fields is shown in Fig. \ref{Model}. The non-interacting QPC model predicts a single peak of $1/T_1$ around $G=0.5\times2e^2/h$ at low temperatures. At a large magnetic field of $B = 5.1$ T, a half-conductance plateau at $G=0.5\times2e^2/h$ is observed, along with a single peak in $1/T_1$ at the conductance riser.
The $0.7$ anomaly is accompanied by an enhanced $g$-factor in finite magnetic fields due to electron-electron interactions [11]. Bauer et al., in Ref. [23], thoroughly explore the $0.7$ structure issue using the Hubbard model, including the mechanism behind the $g$-factor enhancement caused by the Hubbard interaction. They also show that the gate voltage dependence of the spin susceptibility $\chi_{\rm tot}$ peaks near $G \sim 0.7 \times 2e^2/h$, with the peak height increasing as the Hubbard interaction strengthens. In the mean-field approximation of the Hubbard model, taking into account the Zeeman field, the gate voltage dependence of the magnetization $m$ near the center of the QPC exhibits a similar peak structure [48].  In this case, the effective Zeeman energy (spin gap) near the QPC center is 
\begin{equation}  \label{eq:effective-zeeman}
 \tilde{Z_{\rm e}} = Z_e + U m
\end{equation}
with the strength of the Hubbard interaction $U$ and $Um = u\exp{\frac{-(E_F - E_0)^2}{\sigma^2}}$ takes a phenomenological Gaussian function. The coefficients $u$ and $\sigma$ determine the strength of interactions, with $E_0$ being the center of the peak and $E_F$ being the Fermi energy.

We consider  $1/T_1$ for the QPC with the Hubbard interaction given by replacing $Z_{\rm e}$ in Eq.~(\ref{eq:cooper-formula}) with $\tilde{Z_{\rm e}}$ in Eq.~(\ref{eq:effective-zeeman}). When a strong magnetic field is applied, $\tilde{Z_{\rm e}}$ is dominated by the $Z_e$ term, making $U m$ term less visible. For this reason, a single peak structure is seen around $G=0.5 \times 2e^2/h$, as in the non-interacting model, which is consistent with the peak structures for $B = 2.55$ and $5.1$ T. As the magnetic field decreases,  $\tilde{Z_{\rm e}}$ is dominated by the $U m$ term. The spin gap is maximum at the peak of $Um$. Then, the spin-flip rate is suppressed at the peak, and $1/T_1$ is suppressed. As $Um$ moves away from the peak, the spin gap decreases, and the spin-flip rate increases. For this reason, $1/T_1$ increases on both sides of the peak of $Um$, resulting in the double-peak structure of $1/T_1$ as in the experiment at $B=1.7$ T displayed in Fig. 3(d). in the main text.


\begin{thebibliography}{48}%
\makeatletter
\providecommand \@ifxundefined [1]{%
 \@ifx{#1\undefined}
}%
\providecommand \@ifnum [1]{%
 \ifnum #1\expandafter \@firstoftwo
 \else \expandafter \@secondoftwo
 \fi
}%
\providecommand \@ifx [1]{%
 \ifx #1\expandafter \@firstoftwo
 \else \expandafter \@secondoftwo
 \fi
}%
\providecommand \natexlab [1]{#1}%
\providecommand \enquote  [1]{``#1''}%
\providecommand \bibnamefont  [1]{#1}%
\providecommand \bibfnamefont [1]{#1}%
\providecommand \citenamefont [1]{#1}%
\providecommand \href@noop [0]{\@secondoftwo}%
\providecommand \href [0]{\begingroup \@sanitize@url \@href}%
\providecommand \@href[1]{\@@startlink{#1}\@@href}%
\providecommand \@@href[1]{\endgroup#1\@@endlink}%
\providecommand \@sanitize@url [0]{\catcode `\\12\catcode `\$12\catcode `\&12\catcode `\#12\catcode `\^12\catcode `\_12\catcode `\%12\relax}%
\providecommand \@@startlink[1]{}%
\providecommand \@@endlink[0]{}%
\providecommand \url  [0]{\begingroup\@sanitize@url \@url }%
\providecommand \@url [1]{\endgroup\@href {#1}{\urlprefix }}%
\providecommand \urlprefix  [0]{URL }%
\providecommand \Eprint [0]{\href }%
\providecommand \doibase [0]{http://dx.doi.org/}%
\providecommand \selectlanguage [0]{\@gobble}%
\providecommand \bibinfo  [0]{\@secondoftwo}%
\providecommand \bibfield  [0]{\@secondoftwo}%
\providecommand \translation [1]{[#1]}%
\providecommand \BibitemOpen [0]{}%
\providecommand \bibitemStop [0]{}%
\providecommand \bibitemNoStop [0]{.\EOS\space}%
\providecommand \EOS [0]{\spacefactor3000\relax}%
\providecommand \BibitemShut  [1]{\csname bibitem#1\endcsname}%
\let\auto@bib@innerbib\@empty
\bibitem [{\citenamefont {Petersson}\ \emph {et~al.}(2010)\citenamefont {Petersson}, \citenamefont {Petta}, \citenamefont {Lu},\ and\ \citenamefont {Gossard}}]{Petersson2010}%
  \BibitemOpen
  \bibfield  {author} {\bibinfo {author} {\bibfnamefont {K.~D.}\ \bibnamefont {Petersson}}, \bibinfo {author} {\bibfnamefont {J.~R.}\ \bibnamefont {Petta}}, \bibinfo {author} {\bibfnamefont {H.}~\bibnamefont {Lu}}, \ and\ \bibinfo {author} {\bibfnamefont {A.~C.}\ \bibnamefont {Gossard}},\ }\href {\doibase 10.1103/PhysRevLett.105.246804} {\bibfield  {journal} {\bibinfo  {journal} {Phys. Rev. Lett.}\ }\textbf {\bibinfo {volume} {105}},\ \bibinfo {pages} {246804} (\bibinfo {year} {2010})}\BibitemShut {NoStop}%
\bibitem [{\citenamefont {Elzerman}\ \emph {et~al.}(2004)\citenamefont {Elzerman}, \citenamefont {Hanson}, \citenamefont {Willems~van Beveren}, \citenamefont {Witkamp}, \citenamefont {Vandersypen},\ and\ \citenamefont {Kouwenhoven}}]{Elzerman2004}%
  \BibitemOpen
  \bibfield  {author} {\bibinfo {author} {\bibfnamefont {J.~M.}\ \bibnamefont {Elzerman}}, \bibinfo {author} {\bibfnamefont {R.}~\bibnamefont {Hanson}}, \bibinfo {author} {\bibfnamefont {L.~H.}\ \bibnamefont {Willems~van Beveren}}, \bibinfo {author} {\bibfnamefont {B.}~\bibnamefont {Witkamp}}, \bibinfo {author} {\bibfnamefont {L.~M.~K.}\ \bibnamefont {Vandersypen}}, \ and\ \bibinfo {author} {\bibfnamefont {L.~P.}\ \bibnamefont {Kouwenhoven}},\ }\href {\doibase 10.1038/nature02693} {\bibfield  {journal} {\bibinfo  {journal} {Nature}\ }\textbf {\bibinfo {volume} {430}},\ \bibinfo {pages} {431} (\bibinfo {year} {2004})}\BibitemShut {NoStop}%
\bibitem [{\citenamefont {de~Picciotto}\ \emph {et~al.}(1998)\citenamefont {de~Picciotto}, \citenamefont {Reznikov}, \citenamefont {Heiblum}, \citenamefont {Umansky}, \citenamefont {Bunin},\ and\ \citenamefont {Mahalu}}]{DEPICCIOTTO1998395}%
  \BibitemOpen
  \bibfield  {author} {\bibinfo {author} {\bibfnamefont {R.}~\bibnamefont {de~Picciotto}}, \bibinfo {author} {\bibfnamefont {M.}~\bibnamefont {Reznikov}}, \bibinfo {author} {\bibfnamefont {M.}~\bibnamefont {Heiblum}}, \bibinfo {author} {\bibfnamefont {V.}~\bibnamefont {Umansky}}, \bibinfo {author} {\bibfnamefont {G.}~\bibnamefont {Bunin}}, \ and\ \bibinfo {author} {\bibfnamefont {D.}~\bibnamefont {Mahalu}},\ }\href {\doibase https://doi.org/10.1016/S0921-4526(98)00139-2} {\bibfield  {journal} {\bibinfo  {journal} {Physica B: Condensed Matter}\ }\textbf {\bibinfo {volume} {249-251}},\ \bibinfo {pages} {395} (\bibinfo {year} {1998})}\BibitemShut {NoStop}%
\bibitem [{\citenamefont {Ji}\ \emph {et~al.}(2003)\citenamefont {Ji}, \citenamefont {Chung}, \citenamefont {Sprinzak}, \citenamefont {Heiblum}, \citenamefont {Mahalu},\ and\ \citenamefont {Shtrikman}}]{Ji2003}%
  \BibitemOpen
  \bibfield  {author} {\bibinfo {author} {\bibfnamefont {Y.}~\bibnamefont {Ji}}, \bibinfo {author} {\bibfnamefont {Y.}~\bibnamefont {Chung}}, \bibinfo {author} {\bibfnamefont {D.}~\bibnamefont {Sprinzak}}, \bibinfo {author} {\bibfnamefont {M.}~\bibnamefont {Heiblum}}, \bibinfo {author} {\bibfnamefont {D.}~\bibnamefont {Mahalu}}, \ and\ \bibinfo {author} {\bibfnamefont {H.}~\bibnamefont {Shtrikman}},\ }\href {\doibase 10.1038/nature01503} {\bibfield  {journal} {\bibinfo  {journal} {Nature}\ }\textbf {\bibinfo {volume} {422}},\ \bibinfo {pages} {415} (\bibinfo {year} {2003})}\BibitemShut {NoStop}%
\bibitem [{\citenamefont {Baer}\ \emph {et~al.}(2014)\citenamefont {Baer}, \citenamefont {R\"ossler}, \citenamefont {Ihn}, \citenamefont {Ensslin}, \citenamefont {Reichl},\ and\ \citenamefont {Wegscheider}}]{Baer2014}%
  \BibitemOpen
  \bibfield  {author} {\bibinfo {author} {\bibfnamefont {S.}~\bibnamefont {Baer}}, \bibinfo {author} {\bibfnamefont {C.}~\bibnamefont {R\"ossler}}, \bibinfo {author} {\bibfnamefont {T.}~\bibnamefont {Ihn}}, \bibinfo {author} {\bibfnamefont {K.}~\bibnamefont {Ensslin}}, \bibinfo {author} {\bibfnamefont {C.}~\bibnamefont {Reichl}}, \ and\ \bibinfo {author} {\bibfnamefont {W.}~\bibnamefont {Wegscheider}},\ }\href {\doibase 10.1103/PhysRevB.90.075403} {\bibfield  {journal} {\bibinfo  {journal} {Phys. Rev. B}\ }\textbf {\bibinfo {volume} {90}},\ \bibinfo {pages} {075403} (\bibinfo {year} {2014})}\BibitemShut {NoStop}%
\bibitem [{\citenamefont {Banerjee}\ \emph {et~al.}(2018)\citenamefont {Banerjee}, \citenamefont {Heiblum}, \citenamefont {Umansky}, \citenamefont {Feldman}, \citenamefont {Oreg},\ and\ \citenamefont {Stern}}]{Banerjee2018}%
  \BibitemOpen
  \bibfield  {author} {\bibinfo {author} {\bibfnamefont {M.}~\bibnamefont {Banerjee}}, \bibinfo {author} {\bibfnamefont {M.}~\bibnamefont {Heiblum}}, \bibinfo {author} {\bibfnamefont {V.}~\bibnamefont {Umansky}}, \bibinfo {author} {\bibfnamefont {D.~E.}\ \bibnamefont {Feldman}}, \bibinfo {author} {\bibfnamefont {Y.}~\bibnamefont {Oreg}}, \ and\ \bibinfo {author} {\bibfnamefont {A.}~\bibnamefont {Stern}},\ }\href {\doibase 10.1038/s41586-018-0184-1} {\bibfield  {journal} {\bibinfo  {journal} {Nature}\ }\textbf {\bibinfo {volume} {559}},\ \bibinfo {pages} {205} (\bibinfo {year} {2018})}\BibitemShut {NoStop}%
\bibitem [{\citenamefont {Bartolomei}\ \emph {et~al.}(2020)\citenamefont {Bartolomei}, \citenamefont {Kumar}, \citenamefont {Bisognin}, \citenamefont {Marguerite}, \citenamefont {Berroir}, \citenamefont {Bocquillon}, \citenamefont {Plaçais}, \citenamefont {Cavanna}, \citenamefont {Dong}, \citenamefont {Gennser}, \citenamefont {Jin},\ and\ \citenamefont {Fève}}]{Bartolomei2020}%
  \BibitemOpen
  \bibfield  {author} {\bibinfo {author} {\bibfnamefont {H.}~\bibnamefont {Bartolomei}}, \bibinfo {author} {\bibfnamefont {M.}~\bibnamefont {Kumar}}, \bibinfo {author} {\bibfnamefont {R.}~\bibnamefont {Bisognin}}, \bibinfo {author} {\bibfnamefont {A.}~\bibnamefont {Marguerite}}, \bibinfo {author} {\bibfnamefont {J.-M.}\ \bibnamefont {Berroir}}, \bibinfo {author} {\bibfnamefont {E.}~\bibnamefont {Bocquillon}}, \bibinfo {author} {\bibfnamefont {B.}~\bibnamefont {Plaçais}}, \bibinfo {author} {\bibfnamefont {A.}~\bibnamefont {Cavanna}}, \bibinfo {author} {\bibfnamefont {Q.}~\bibnamefont {Dong}}, \bibinfo {author} {\bibfnamefont {U.}~\bibnamefont {Gennser}}, \bibinfo {author} {\bibfnamefont {Y.}~\bibnamefont {Jin}}, \ and\ \bibinfo {author} {\bibfnamefont {G.}~\bibnamefont {Fève}},\ }\href {\doibase 10.1126/science.aaz5601} {\bibfield  {journal} {\bibinfo  {journal} {Science}\ }\textbf {\bibinfo {volume} {368}},\ \bibinfo {pages} {173} (\bibinfo {year} {2020})},\ \Eprint
  {http://arxiv.org/abs/https://www.science.org/doi/pdf/10.1126/science.aaz5601} {https://www.science.org/doi/pdf/10.1126/science.aaz5601} \BibitemShut {NoStop}%
\bibitem [{\citenamefont {Nakamura}\ \emph {et~al.}(2019)\citenamefont {Nakamura}, \citenamefont {Fallahi}, \citenamefont {Sahasrabudhe}, \citenamefont {Rahman}, \citenamefont {Liang}, \citenamefont {Gardner},\ and\ \citenamefont {Manfra}}]{Nakamura2019}%
  \BibitemOpen
  \bibfield  {author} {\bibinfo {author} {\bibfnamefont {J.}~\bibnamefont {Nakamura}}, \bibinfo {author} {\bibfnamefont {S.}~\bibnamefont {Fallahi}}, \bibinfo {author} {\bibfnamefont {H.}~\bibnamefont {Sahasrabudhe}}, \bibinfo {author} {\bibfnamefont {R.}~\bibnamefont {Rahman}}, \bibinfo {author} {\bibfnamefont {S.}~\bibnamefont {Liang}}, \bibinfo {author} {\bibfnamefont {G.~C.}\ \bibnamefont {Gardner}}, \ and\ \bibinfo {author} {\bibfnamefont {M.~J.}\ \bibnamefont {Manfra}},\ }\href {\doibase 10.1038/s41567-019-0441-8} {\bibfield  {journal} {\bibinfo  {journal} {Nature Physics}\ }\textbf {\bibinfo {volume} {15}},\ \bibinfo {pages} {563} (\bibinfo {year} {2019})}\BibitemShut {NoStop}%
\bibitem [{\citenamefont {Nakamura}\ \emph {et~al.}(2020)\citenamefont {Nakamura}, \citenamefont {Liang}, \citenamefont {Gardner},\ and\ \citenamefont {Manfra}}]{Nakamura2020}%
  \BibitemOpen
  \bibfield  {author} {\bibinfo {author} {\bibfnamefont {J.}~\bibnamefont {Nakamura}}, \bibinfo {author} {\bibfnamefont {S.}~\bibnamefont {Liang}}, \bibinfo {author} {\bibfnamefont {G.~C.}\ \bibnamefont {Gardner}}, \ and\ \bibinfo {author} {\bibfnamefont {M.~J.}\ \bibnamefont {Manfra}},\ }\href {\doibase 10.1038/s41567-020-1019-1} {\bibfield  {journal} {\bibinfo  {journal} {Nature Physics}\ }\textbf {\bibinfo {volume} {16}},\ \bibinfo {pages} {931} (\bibinfo {year} {2020})}\BibitemShut {NoStop}%
\bibitem [{\citenamefont {Thomas}\ \emph {et~al.}(1996)\citenamefont {Thomas}, \citenamefont {Nicholls}, \citenamefont {Simmons}, \citenamefont {Pepper}, \citenamefont {Mace},\ and\ \citenamefont {Ritchie}}]{Thomas}%
  \BibitemOpen
  \bibfield  {author} {\bibinfo {author} {\bibfnamefont {K.~J.}\ \bibnamefont {Thomas}}, \bibinfo {author} {\bibfnamefont {J.~T.}\ \bibnamefont {Nicholls}}, \bibinfo {author} {\bibfnamefont {M.~Y.}\ \bibnamefont {Simmons}}, \bibinfo {author} {\bibfnamefont {M.}~\bibnamefont {Pepper}}, \bibinfo {author} {\bibfnamefont {D.~R.}\ \bibnamefont {Mace}}, \ and\ \bibinfo {author} {\bibfnamefont {D.~A.}\ \bibnamefont {Ritchie}},\ }\href {\doibase 10.1103/PhysRevLett.77.135} {\bibfield  {journal} {\bibinfo  {journal} {Phys. Rev. Lett.}\ }\textbf {\bibinfo {volume} {77}},\ \bibinfo {pages} {135} (\bibinfo {year} {1996})}\BibitemShut {NoStop}%
\bibitem [{\citenamefont {Thomas}\ \emph {et~al.}(1998)\citenamefont {Thomas}, \citenamefont {Nicholls}, \citenamefont {Appleyard}, \citenamefont {Simmons}, \citenamefont {Pepper}, \citenamefont {Mace}, \citenamefont {Tribe},\ and\ \citenamefont {Ritchie}}]{Thomas98}%
  \BibitemOpen
  \bibfield  {author} {\bibinfo {author} {\bibfnamefont {K.~J.}\ \bibnamefont {Thomas}}, \bibinfo {author} {\bibfnamefont {J.~T.}\ \bibnamefont {Nicholls}}, \bibinfo {author} {\bibfnamefont {N.~J.}\ \bibnamefont {Appleyard}}, \bibinfo {author} {\bibfnamefont {M.~Y.}\ \bibnamefont {Simmons}}, \bibinfo {author} {\bibfnamefont {M.}~\bibnamefont {Pepper}}, \bibinfo {author} {\bibfnamefont {D.~R.}\ \bibnamefont {Mace}}, \bibinfo {author} {\bibfnamefont {W.~R.}\ \bibnamefont {Tribe}}, \ and\ \bibinfo {author} {\bibfnamefont {D.~A.}\ \bibnamefont {Ritchie}},\ }\href {\doibase 10.1103/PhysRevB.58.4846} {\bibfield  {journal} {\bibinfo  {journal} {Phys. Rev. B}\ }\textbf {\bibinfo {volume} {58}},\ \bibinfo {pages} {4846} (\bibinfo {year} {1998})}\BibitemShut {NoStop}%
\bibitem [{\citenamefont {Nuttinck}\ \emph {et~al.}(2000)\citenamefont {Nuttinck}, \citenamefont {Hashimoto}, \citenamefont {Miyashita}, \citenamefont {Saku}, \citenamefont {Yamamoto},\ and\ \citenamefont {Hirayama}}]{Sebastien2000}%
  \BibitemOpen
  \bibfield  {author} {\bibinfo {author} {\bibfnamefont {S.~N.~S.}\ \bibnamefont {Nuttinck}}, \bibinfo {author} {\bibfnamefont {K.}~\bibnamefont {Hashimoto}}, \bibinfo {author} {\bibfnamefont {S.}~\bibnamefont {Miyashita}}, \bibinfo {author} {\bibfnamefont {T.}~\bibnamefont {Saku}}, \bibinfo {author} {\bibfnamefont {Y.}~\bibnamefont {Yamamoto}}, \ and\ \bibinfo {author} {\bibfnamefont {Y.}~\bibnamefont {Hirayama}},\ }\href {\doibase 10.1143/JJAP.39.L655} {\bibfield  {journal} {\bibinfo  {journal} {Japanese Journal of Applied Physics}\ }\textbf {\bibinfo {volume} {39}},\ \bibinfo {pages} {L655} (\bibinfo {year} {2000})}\BibitemShut {NoStop}%
\bibitem [{\citenamefont {Kristensen}\ \emph {et~al.}(2000)\citenamefont {Kristensen}, \citenamefont {Bruus}, \citenamefont {Hansen}, \citenamefont {Jensen}, \citenamefont {Lindelof}, \citenamefont {Marckmann}, \citenamefont {Nyg\aa{}rd}, \citenamefont {S\o{}rensen}, \citenamefont {Beuscher}, \citenamefont {Forchel},\ and\ \citenamefont {Michel}}]{Kristensen2000}%
  \BibitemOpen
  \bibfield  {author} {\bibinfo {author} {\bibfnamefont {A.}~\bibnamefont {Kristensen}}, \bibinfo {author} {\bibfnamefont {H.}~\bibnamefont {Bruus}}, \bibinfo {author} {\bibfnamefont {A.~E.}\ \bibnamefont {Hansen}}, \bibinfo {author} {\bibfnamefont {J.~B.}\ \bibnamefont {Jensen}}, \bibinfo {author} {\bibfnamefont {P.~E.}\ \bibnamefont {Lindelof}}, \bibinfo {author} {\bibfnamefont {C.~J.}\ \bibnamefont {Marckmann}}, \bibinfo {author} {\bibfnamefont {J.}~\bibnamefont {Nyg\aa{}rd}}, \bibinfo {author} {\bibfnamefont {C.~B.}\ \bibnamefont {S\o{}rensen}}, \bibinfo {author} {\bibfnamefont {F.}~\bibnamefont {Beuscher}}, \bibinfo {author} {\bibfnamefont {A.}~\bibnamefont {Forchel}}, \ and\ \bibinfo {author} {\bibfnamefont {M.}~\bibnamefont {Michel}},\ }\href {\doibase 10.1103/PhysRevB.62.10950} {\bibfield  {journal} {\bibinfo  {journal} {Phys. Rev. B}\ }\textbf {\bibinfo {volume} {62}},\ \bibinfo {pages} {10950} (\bibinfo {year} {2000})}\BibitemShut {NoStop}%
\bibitem [{\citenamefont {Reilly}\ \emph {et~al.}(2002)\citenamefont {Reilly}, \citenamefont {Buehler}, \citenamefont {O'Brien}, \citenamefont {Hamilton}, \citenamefont {Dzurak}, \citenamefont {Clark}, \citenamefont {Kane}, \citenamefont {Pfeiffer},\ and\ \citenamefont {West}}]{Reilly2002}%
  \BibitemOpen
  \bibfield  {author} {\bibinfo {author} {\bibfnamefont {D.~J.}\ \bibnamefont {Reilly}}, \bibinfo {author} {\bibfnamefont {T.~M.}\ \bibnamefont {Buehler}}, \bibinfo {author} {\bibfnamefont {J.~L.}\ \bibnamefont {O'Brien}}, \bibinfo {author} {\bibfnamefont {A.~R.}\ \bibnamefont {Hamilton}}, \bibinfo {author} {\bibfnamefont {A.~S.}\ \bibnamefont {Dzurak}}, \bibinfo {author} {\bibfnamefont {R.~G.}\ \bibnamefont {Clark}}, \bibinfo {author} {\bibfnamefont {B.~E.}\ \bibnamefont {Kane}}, \bibinfo {author} {\bibfnamefont {L.~N.}\ \bibnamefont {Pfeiffer}}, \ and\ \bibinfo {author} {\bibfnamefont {K.~W.}\ \bibnamefont {West}},\ }\href {\doibase 10.1103/PhysRevLett.89.246801} {\bibfield  {journal} {\bibinfo  {journal} {Phys. Rev. Lett.}\ }\textbf {\bibinfo {volume} {89}},\ \bibinfo {pages} {246801} (\bibinfo {year} {2002})}\BibitemShut {NoStop}%
\bibitem [{\citenamefont {Cronenwett}\ \emph {et~al.}(2002)\citenamefont {Cronenwett}, \citenamefont {Lynch}, \citenamefont {Goldhaber-Gordon}, \citenamefont {Kouwenhoven}, \citenamefont {Marcus}, \citenamefont {Hirose}, \citenamefont {Wingreen},\ and\ \citenamefont {Umansky}}]{Cronenwett}%
  \BibitemOpen
  \bibfield  {author} {\bibinfo {author} {\bibfnamefont {S.~M.}\ \bibnamefont {Cronenwett}}, \bibinfo {author} {\bibfnamefont {H.~J.}\ \bibnamefont {Lynch}}, \bibinfo {author} {\bibfnamefont {D.}~\bibnamefont {Goldhaber-Gordon}}, \bibinfo {author} {\bibfnamefont {L.~P.}\ \bibnamefont {Kouwenhoven}}, \bibinfo {author} {\bibfnamefont {C.~M.}\ \bibnamefont {Marcus}}, \bibinfo {author} {\bibfnamefont {K.}~\bibnamefont {Hirose}}, \bibinfo {author} {\bibfnamefont {N.~S.}\ \bibnamefont {Wingreen}}, \ and\ \bibinfo {author} {\bibfnamefont {V.}~\bibnamefont {Umansky}},\ }\href {\doibase 10.1103/PhysRevLett.88.226805} {\bibfield  {journal} {\bibinfo  {journal} {Phys. Rev. Lett.}\ }\textbf {\bibinfo {volume} {88}},\ \bibinfo {pages} {226805} (\bibinfo {year} {2002})}\BibitemShut {NoStop}%
\bibitem [{\citenamefont {DiCarlo}\ \emph {et~al.}(2006)\citenamefont {DiCarlo}, \citenamefont {Zhang}, \citenamefont {McClure}, \citenamefont {Reilly}, \citenamefont {Marcus}, \citenamefont {Pfeiffer},\ and\ \citenamefont {West}}]{DiCarlo2006}%
  \BibitemOpen
  \bibfield  {author} {\bibinfo {author} {\bibfnamefont {L.}~\bibnamefont {DiCarlo}}, \bibinfo {author} {\bibfnamefont {Y.}~\bibnamefont {Zhang}}, \bibinfo {author} {\bibfnamefont {D.~T.}\ \bibnamefont {McClure}}, \bibinfo {author} {\bibfnamefont {D.~J.}\ \bibnamefont {Reilly}}, \bibinfo {author} {\bibfnamefont {C.~M.}\ \bibnamefont {Marcus}}, \bibinfo {author} {\bibfnamefont {L.~N.}\ \bibnamefont {Pfeiffer}}, \ and\ \bibinfo {author} {\bibfnamefont {K.~W.}\ \bibnamefont {West}},\ }\href {\doibase 10.1103/PhysRevLett.97.036810} {\bibfield  {journal} {\bibinfo  {journal} {Phys. Rev. Lett.}\ }\textbf {\bibinfo {volume} {97}},\ \bibinfo {pages} {036810} (\bibinfo {year} {2006})}\BibitemShut {NoStop}%
\bibitem [{\citenamefont {Chung}\ \emph {et~al.}(2007)\citenamefont {Chung}, \citenamefont {Jo}, \citenamefont {Chang}, \citenamefont {Lee}, \citenamefont {Zaffalon}, \citenamefont {Umansky},\ and\ \citenamefont {Heiblum}}]{Chung2007}%
  \BibitemOpen
  \bibfield  {author} {\bibinfo {author} {\bibfnamefont {Y.}~\bibnamefont {Chung}}, \bibinfo {author} {\bibfnamefont {S.}~\bibnamefont {Jo}}, \bibinfo {author} {\bibfnamefont {D.-I.}\ \bibnamefont {Chang}}, \bibinfo {author} {\bibfnamefont {H.-J.}\ \bibnamefont {Lee}}, \bibinfo {author} {\bibfnamefont {M.}~\bibnamefont {Zaffalon}}, \bibinfo {author} {\bibfnamefont {V.}~\bibnamefont {Umansky}}, \ and\ \bibinfo {author} {\bibfnamefont {M.}~\bibnamefont {Heiblum}},\ }\href {\doibase 10.1103/PhysRevB.76.035316} {\bibfield  {journal} {\bibinfo  {journal} {Phys. Rev. B}\ }\textbf {\bibinfo {volume} {76}},\ \bibinfo {pages} {035316} (\bibinfo {year} {2007})}\BibitemShut {NoStop}%
\bibitem [{\citenamefont {Koop}\ \emph {et~al.}(2007)\citenamefont {Koop}, \citenamefont {Lerescu}, \citenamefont {Liu}, \citenamefont {van Wees}, \citenamefont {Reuter}, \citenamefont {Wieck},\ and\ \citenamefont {van~der Wal}}]{Koop2007}%
  \BibitemOpen
  \bibfield  {author} {\bibinfo {author} {\bibfnamefont {E.~J.}\ \bibnamefont {Koop}}, \bibinfo {author} {\bibfnamefont {A.~I.}\ \bibnamefont {Lerescu}}, \bibinfo {author} {\bibfnamefont {J.}~\bibnamefont {Liu}}, \bibinfo {author} {\bibfnamefont {B.~J.}\ \bibnamefont {van Wees}}, \bibinfo {author} {\bibfnamefont {D.}~\bibnamefont {Reuter}}, \bibinfo {author} {\bibfnamefont {A.~D.}\ \bibnamefont {Wieck}}, \ and\ \bibinfo {author} {\bibfnamefont {C.~H.}\ \bibnamefont {van~der Wal}},\ }\href {\doibase 10.1007/s10948-007-0289-5} {\bibfield  {journal} {\bibinfo  {journal} {Journal of Superconductivity and Novel Magnetism}\ }\textbf {\bibinfo {volume} {20}},\ \bibinfo {pages} {433} (\bibinfo {year} {2007})}\BibitemShut {NoStop}%
\bibitem [{\citenamefont {Nakamura}\ \emph {et~al.}(2009)\citenamefont {Nakamura}, \citenamefont {Hashisaka}, \citenamefont {Yamauchi}, \citenamefont {Kasai}, \citenamefont {Ono},\ and\ \citenamefont {Kobayashi}}]{SNakamura2009}%
  \BibitemOpen
  \bibfield  {author} {\bibinfo {author} {\bibfnamefont {S.}~\bibnamefont {Nakamura}}, \bibinfo {author} {\bibfnamefont {M.}~\bibnamefont {Hashisaka}}, \bibinfo {author} {\bibfnamefont {Y.}~\bibnamefont {Yamauchi}}, \bibinfo {author} {\bibfnamefont {S.}~\bibnamefont {Kasai}}, \bibinfo {author} {\bibfnamefont {T.}~\bibnamefont {Ono}}, \ and\ \bibinfo {author} {\bibfnamefont {K.}~\bibnamefont {Kobayashi}},\ }\href {\doibase 10.1103/PhysRevB.79.201308} {\bibfield  {journal} {\bibinfo  {journal} {Phys. Rev. B}\ }\textbf {\bibinfo {volume} {79}},\ \bibinfo {pages} {201308} (\bibinfo {year} {2009})}\BibitemShut {NoStop}%
\bibitem [{\citenamefont {Komijani}\ \emph {et~al.}(2010)\citenamefont {Komijani}, \citenamefont {Csontos}, \citenamefont {Shorubalko}, \citenamefont {Ihn}, \citenamefont {Ensslin}, \citenamefont {Meir}, \citenamefont {Reuter},\ and\ \citenamefont {Wieck}}]{Komijani2010}%
  \BibitemOpen
  \bibfield  {author} {\bibinfo {author} {\bibfnamefont {Y.}~\bibnamefont {Komijani}}, \bibinfo {author} {\bibfnamefont {M.}~\bibnamefont {Csontos}}, \bibinfo {author} {\bibfnamefont {I.}~\bibnamefont {Shorubalko}}, \bibinfo {author} {\bibfnamefont {T.}~\bibnamefont {Ihn}}, \bibinfo {author} {\bibfnamefont {K.}~\bibnamefont {Ensslin}}, \bibinfo {author} {\bibfnamefont {Y.}~\bibnamefont {Meir}}, \bibinfo {author} {\bibfnamefont {D.}~\bibnamefont {Reuter}}, \ and\ \bibinfo {author} {\bibfnamefont {A.~D.}\ \bibnamefont {Wieck}},\ }\href {\doibase 10.1209/0295-5075/91/67010} {\bibfield  {journal} {\bibinfo  {journal} {Europhysics Letters}\ }\textbf {\bibinfo {volume} {91}},\ \bibinfo {pages} {67010} (\bibinfo {year} {2010})}\BibitemShut {NoStop}%
\bibitem [{\citenamefont {Smith}\ \emph {et~al.}(2011)\citenamefont {Smith}, \citenamefont {Hamilton}, \citenamefont {Thomas}, \citenamefont {Pepper}, \citenamefont {Farrer}, \citenamefont {Griffiths}, \citenamefont {Jones},\ and\ \citenamefont {Ritchie}}]{Smith2011}%
  \BibitemOpen
  \bibfield  {author} {\bibinfo {author} {\bibfnamefont {L.~W.}\ \bibnamefont {Smith}}, \bibinfo {author} {\bibfnamefont {A.~R.}\ \bibnamefont {Hamilton}}, \bibinfo {author} {\bibfnamefont {K.~J.}\ \bibnamefont {Thomas}}, \bibinfo {author} {\bibfnamefont {M.}~\bibnamefont {Pepper}}, \bibinfo {author} {\bibfnamefont {I.}~\bibnamefont {Farrer}}, \bibinfo {author} {\bibfnamefont {J.~P.}\ \bibnamefont {Griffiths}}, \bibinfo {author} {\bibfnamefont {G.~A.~C.}\ \bibnamefont {Jones}}, \ and\ \bibinfo {author} {\bibfnamefont {D.~A.}\ \bibnamefont {Ritchie}},\ }\href {\doibase 10.1103/PhysRevLett.107.126801} {\bibfield  {journal} {\bibinfo  {journal} {Phys. Rev. Lett.}\ }\textbf {\bibinfo {volume} {107}},\ \bibinfo {pages} {126801} (\bibinfo {year} {2011})}\BibitemShut {NoStop}%
\bibitem [{\citenamefont {Iqbal}\ \emph {et~al.}(2013)\citenamefont {Iqbal}, \citenamefont {Levy}, \citenamefont {Koop}, \citenamefont {Dekker}, \citenamefont {de~Jong}, \citenamefont {van~der Velde}, \citenamefont {Reuter}, \citenamefont {Wieck}, \citenamefont {Aguado}, \citenamefont {Meir},\ and\ \citenamefont {van~der Wal}}]{Iqbal2013}%
  \BibitemOpen
  \bibfield  {author} {\bibinfo {author} {\bibfnamefont {M.~J.}\ \bibnamefont {Iqbal}}, \bibinfo {author} {\bibfnamefont {R.}~\bibnamefont {Levy}}, \bibinfo {author} {\bibfnamefont {E.~J.}\ \bibnamefont {Koop}}, \bibinfo {author} {\bibfnamefont {J.~B.}\ \bibnamefont {Dekker}}, \bibinfo {author} {\bibfnamefont {J.~P.}\ \bibnamefont {de~Jong}}, \bibinfo {author} {\bibfnamefont {J.~H.~M.}\ \bibnamefont {van~der Velde}}, \bibinfo {author} {\bibfnamefont {D.}~\bibnamefont {Reuter}}, \bibinfo {author} {\bibfnamefont {A.~D.}\ \bibnamefont {Wieck}}, \bibinfo {author} {\bibfnamefont {R.}~\bibnamefont {Aguado}}, \bibinfo {author} {\bibfnamefont {Y.}~\bibnamefont {Meir}}, \ and\ \bibinfo {author} {\bibfnamefont {C.~H.}\ \bibnamefont {van~der Wal}},\ }\href {\doibase 10.1038/nature12491} {\bibfield  {journal} {\bibinfo  {journal} {Nature}\ }\textbf {\bibinfo {volume} {501}},\ \bibinfo {pages} {79} (\bibinfo {year} {2013})}\BibitemShut {NoStop}%
\bibitem [{\citenamefont {Bauer}\ \emph {et~al.}(2013)\citenamefont {Bauer}, \citenamefont {Heyder}, \citenamefont {Schubert}, \citenamefont {Borowsky}, \citenamefont {Taubert}, \citenamefont {Bruognolo}, \citenamefont {Schuh}, \citenamefont {Wegscheider}, \citenamefont {von Delft},\ and\ \citenamefont {Ludwig}}]{Bauer2013}%
  \BibitemOpen
  \bibfield  {author} {\bibinfo {author} {\bibfnamefont {F.}~\bibnamefont {Bauer}}, \bibinfo {author} {\bibfnamefont {J.}~\bibnamefont {Heyder}}, \bibinfo {author} {\bibfnamefont {E.}~\bibnamefont {Schubert}}, \bibinfo {author} {\bibfnamefont {D.}~\bibnamefont {Borowsky}}, \bibinfo {author} {\bibfnamefont {D.}~\bibnamefont {Taubert}}, \bibinfo {author} {\bibfnamefont {B.}~\bibnamefont {Bruognolo}}, \bibinfo {author} {\bibfnamefont {D.}~\bibnamefont {Schuh}}, \bibinfo {author} {\bibfnamefont {W.}~\bibnamefont {Wegscheider}}, \bibinfo {author} {\bibfnamefont {J.}~\bibnamefont {von Delft}}, \ and\ \bibinfo {author} {\bibfnamefont {S.}~\bibnamefont {Ludwig}},\ }\href {\doibase 10.1038/nature12421} {\bibfield  {journal} {\bibinfo  {journal} {Nature}\ }\textbf {\bibinfo {volume} {501}},\ \bibinfo {pages} {73} (\bibinfo {year} {2013})}\BibitemShut {NoStop}%
\bibitem [{\citenamefont {Smith}\ \emph {et~al.}(2014)\citenamefont {Smith}, \citenamefont {Al-Taie}, \citenamefont {Sfigakis}, \citenamefont {See}, \citenamefont {Lesage}, \citenamefont {Xu}, \citenamefont {Griffiths}, \citenamefont {Beere}, \citenamefont {Jones}, \citenamefont {Ritchie}, \citenamefont {Kelly},\ and\ \citenamefont {Smith}}]{Smith2014}%
  \BibitemOpen
  \bibfield  {author} {\bibinfo {author} {\bibfnamefont {L.~W.}\ \bibnamefont {Smith}}, \bibinfo {author} {\bibfnamefont {H.}~\bibnamefont {Al-Taie}}, \bibinfo {author} {\bibfnamefont {F.}~\bibnamefont {Sfigakis}}, \bibinfo {author} {\bibfnamefont {P.}~\bibnamefont {See}}, \bibinfo {author} {\bibfnamefont {A.~A.~J.}\ \bibnamefont {Lesage}}, \bibinfo {author} {\bibfnamefont {B.}~\bibnamefont {Xu}}, \bibinfo {author} {\bibfnamefont {J.~P.}\ \bibnamefont {Griffiths}}, \bibinfo {author} {\bibfnamefont {H.~E.}\ \bibnamefont {Beere}}, \bibinfo {author} {\bibfnamefont {G.~A.~C.}\ \bibnamefont {Jones}}, \bibinfo {author} {\bibfnamefont {D.~A.}\ \bibnamefont {Ritchie}}, \bibinfo {author} {\bibfnamefont {M.~J.}\ \bibnamefont {Kelly}}, \ and\ \bibinfo {author} {\bibfnamefont {C.~G.}\ \bibnamefont {Smith}},\ }\href {\doibase 10.1103/PhysRevB.90.045426} {\bibfield  {journal} {\bibinfo  {journal} {Phys. Rev. B}\ }\textbf {\bibinfo {volume} {90}},\ \bibinfo {pages} {045426} (\bibinfo {year} {2014})}\BibitemShut {NoStop}%
\bibitem [{\citenamefont {Smith}\ \emph {et~al.}(2015)\citenamefont {Smith}, \citenamefont {Al-Taie}, \citenamefont {Lesage}, \citenamefont {Sfigakis}, \citenamefont {See}, \citenamefont {Griffiths}, \citenamefont {Beere}, \citenamefont {Jones}, \citenamefont {Ritchie}, \citenamefont {Hamilton}, \citenamefont {Kelly},\ and\ \citenamefont {Smith}}]{Smith2015}%
  \BibitemOpen
  \bibfield  {author} {\bibinfo {author} {\bibfnamefont {L.~W.}\ \bibnamefont {Smith}}, \bibinfo {author} {\bibfnamefont {H.}~\bibnamefont {Al-Taie}}, \bibinfo {author} {\bibfnamefont {A.~A.~J.}\ \bibnamefont {Lesage}}, \bibinfo {author} {\bibfnamefont {F.}~\bibnamefont {Sfigakis}}, \bibinfo {author} {\bibfnamefont {P.}~\bibnamefont {See}}, \bibinfo {author} {\bibfnamefont {J.~P.}\ \bibnamefont {Griffiths}}, \bibinfo {author} {\bibfnamefont {H.~E.}\ \bibnamefont {Beere}}, \bibinfo {author} {\bibfnamefont {G.~A.~C.}\ \bibnamefont {Jones}}, \bibinfo {author} {\bibfnamefont {D.~A.}\ \bibnamefont {Ritchie}}, \bibinfo {author} {\bibfnamefont {A.~R.}\ \bibnamefont {Hamilton}}, \bibinfo {author} {\bibfnamefont {M.~J.}\ \bibnamefont {Kelly}}, \ and\ \bibinfo {author} {\bibfnamefont {C.~G.}\ \bibnamefont {Smith}},\ }\href {\doibase 10.1103/PhysRevB.91.235402} {\bibfield  {journal} {\bibinfo  {journal} {Phys. Rev. B}\ }\textbf {\bibinfo {volume} {91}},\ \bibinfo {pages} {235402} (\bibinfo {year} {2015})}\BibitemShut
  {NoStop}%
\bibitem [{\citenamefont {Smith}\ \emph {et~al.}(2016)\citenamefont {Smith}, \citenamefont {Al-Taie}, \citenamefont {Lesage}, \citenamefont {Thomas}, \citenamefont {Sfigakis}, \citenamefont {See}, \citenamefont {Griffiths}, \citenamefont {Farrer}, \citenamefont {Jones}, \citenamefont {Ritchie}, \citenamefont {Kelly},\ and\ \citenamefont {Smith}}]{Smith2016}%
  \BibitemOpen
  \bibfield  {author} {\bibinfo {author} {\bibfnamefont {L.~W.}\ \bibnamefont {Smith}}, \bibinfo {author} {\bibfnamefont {H.}~\bibnamefont {Al-Taie}}, \bibinfo {author} {\bibfnamefont {A.~A.~J.}\ \bibnamefont {Lesage}}, \bibinfo {author} {\bibfnamefont {K.~J.}\ \bibnamefont {Thomas}}, \bibinfo {author} {\bibfnamefont {F.}~\bibnamefont {Sfigakis}}, \bibinfo {author} {\bibfnamefont {P.}~\bibnamefont {See}}, \bibinfo {author} {\bibfnamefont {J.~P.}\ \bibnamefont {Griffiths}}, \bibinfo {author} {\bibfnamefont {I.}~\bibnamefont {Farrer}}, \bibinfo {author} {\bibfnamefont {G.~A.~C.}\ \bibnamefont {Jones}}, \bibinfo {author} {\bibfnamefont {D.~A.}\ \bibnamefont {Ritchie}}, \bibinfo {author} {\bibfnamefont {M.~J.}\ \bibnamefont {Kelly}}, \ and\ \bibinfo {author} {\bibfnamefont {C.~G.}\ \bibnamefont {Smith}},\ }\href {\doibase 10.1103/PhysRevApplied.5.044015} {\bibfield  {journal} {\bibinfo  {journal} {Phys. Rev. Appl.}\ }\textbf {\bibinfo {volume} {5}},\ \bibinfo {pages} {044015} (\bibinfo {year} {2016})}\BibitemShut
  {NoStop}%
\bibitem [{\citenamefont {Ma}\ \emph {et~al.}(2024)\citenamefont {Ma}, \citenamefont {Delfanazari}, \citenamefont {Puddy}, \citenamefont {Li}, \citenamefont {Cao}, \citenamefont {Yi}, \citenamefont {Griffiths}, \citenamefont {Beere}, \citenamefont {Ritchie}, \citenamefont {Kelly},\ and\ \citenamefont {Smith}}]{MA2024}%
  \BibitemOpen
  \bibfield  {author} {\bibinfo {author} {\bibfnamefont {P.}~\bibnamefont {Ma}}, \bibinfo {author} {\bibfnamefont {K.}~\bibnamefont {Delfanazari}}, \bibinfo {author} {\bibfnamefont {R.~K.}\ \bibnamefont {Puddy}}, \bibinfo {author} {\bibfnamefont {J.}~\bibnamefont {Li}}, \bibinfo {author} {\bibfnamefont {M.}~\bibnamefont {Cao}}, \bibinfo {author} {\bibfnamefont {T.}~\bibnamefont {Yi}}, \bibinfo {author} {\bibfnamefont {J.~P.}\ \bibnamefont {Griffiths}}, \bibinfo {author} {\bibfnamefont {H.~E.}\ \bibnamefont {Beere}}, \bibinfo {author} {\bibfnamefont {D.~A.}\ \bibnamefont {Ritchie}}, \bibinfo {author} {\bibfnamefont {M.~J.}\ \bibnamefont {Kelly}}, \ and\ \bibinfo {author} {\bibfnamefont {C.~G.}\ \bibnamefont {Smith}},\ }\href {\doibase https://doi.org/10.1016/j.chip.2024.100095} {\bibfield  {journal} {\bibinfo  {journal} {Chip}\ ,\ \bibinfo {pages} {100095}} (\bibinfo {year} {2024})}\BibitemShut {NoStop}%
\bibitem [{\citenamefont {Aono}\ \emph {et~al.}(2020)\citenamefont {Aono}, \citenamefont {Takahashi}, \citenamefont {Fauzi},\ and\ \citenamefont {Hirayama}}]{Aono2020}%
  \BibitemOpen
  \bibfield  {author} {\bibinfo {author} {\bibfnamefont {T.}~\bibnamefont {Aono}}, \bibinfo {author} {\bibfnamefont {M.}~\bibnamefont {Takahashi}}, \bibinfo {author} {\bibfnamefont {M.~H.}\ \bibnamefont {Fauzi}}, \ and\ \bibinfo {author} {\bibfnamefont {Y.}~\bibnamefont {Hirayama}},\ }\href {\doibase 10.1103/PhysRevB.102.045305} {\bibfield  {journal} {\bibinfo  {journal} {Phys. Rev. B}\ }\textbf {\bibinfo {volume} {102}},\ \bibinfo {pages} {045305} (\bibinfo {year} {2020})}\BibitemShut {NoStop}%
\bibitem [{\citenamefont {Schimmel}\ \emph {et~al.}(2017)\citenamefont {Schimmel}, \citenamefont {Bruognolo},\ and\ \citenamefont {von Delft}}]{Schimmel2017}%
  \BibitemOpen
  \bibfield  {author} {\bibinfo {author} {\bibfnamefont {D.~H.}\ \bibnamefont {Schimmel}}, \bibinfo {author} {\bibfnamefont {B.}~\bibnamefont {Bruognolo}}, \ and\ \bibinfo {author} {\bibfnamefont {J.}~\bibnamefont {von Delft}},\ }\href {\doibase 10.1103/PhysRevLett.119.196401} {\bibfield  {journal} {\bibinfo  {journal} {Phys. Rev. Lett.}\ }\textbf {\bibinfo {volume} {119}},\ \bibinfo {pages} {196401} (\bibinfo {year} {2017})}\BibitemShut {NoStop}%
\bibitem [{\citenamefont {Ensslin}(2006)}]{Ensslin2006}%
  \BibitemOpen
  \bibfield  {author} {\bibinfo {author} {\bibfnamefont {K.}~\bibnamefont {Ensslin}},\ }\href {\doibase 10.1038/nphys399} {\bibfield  {journal} {\bibinfo  {journal} {Nature Physics}\ }\textbf {\bibinfo {volume} {2}},\ \bibinfo {pages} {587} (\bibinfo {year} {2006})}\BibitemShut {NoStop}%
\bibitem [{\citenamefont {Hirayama}\ \emph {et~al.}(2009)\citenamefont {Hirayama}, \citenamefont {Yusa}, \citenamefont {Hashimoto}, \citenamefont {Kumada}, \citenamefont {Ota},\ and\ \citenamefont {Muraki}}]{Hirayama_2009}%
  \BibitemOpen
  \bibfield  {author} {\bibinfo {author} {\bibfnamefont {Y.}~\bibnamefont {Hirayama}}, \bibinfo {author} {\bibfnamefont {G.}~\bibnamefont {Yusa}}, \bibinfo {author} {\bibfnamefont {K.}~\bibnamefont {Hashimoto}}, \bibinfo {author} {\bibfnamefont {N.}~\bibnamefont {Kumada}}, \bibinfo {author} {\bibfnamefont {T.}~\bibnamefont {Ota}}, \ and\ \bibinfo {author} {\bibfnamefont {K.}~\bibnamefont {Muraki}},\ }\href {\doibase 10.1088/0268-1242/24/2/023001} {\bibfield  {journal} {\bibinfo  {journal} {Semiconductor Science and Technology}\ }\textbf {\bibinfo {volume} {24}},\ \bibinfo {pages} {023001} (\bibinfo {year} {2009})}\BibitemShut {NoStop}%
\bibitem [{\citenamefont {Fauzi}\ and\ \citenamefont {Hirayama}(2022)}]{Fauzi2022}%
  \BibitemOpen
  \bibfield  {author} {\bibinfo {author} {\bibfnamefont {M.~H.}\ \bibnamefont {Fauzi}}\ and\ \bibinfo {author} {\bibfnamefont {Y.}~\bibnamefont {Hirayama}},\ }\enquote {\bibinfo {title} {Hyperfine-mediated transport in a one-dimensional channel},}\ in\ \href {\doibase 10.1007/978-981-19-1201-6_12} {\emph {\bibinfo {booktitle} {Quantum Hybrid Electronics and Materials}}},\ \bibinfo {editor} {edited by\ \bibinfo {editor} {\bibfnamefont {Y.}~\bibnamefont {Hirayama}}, \bibinfo {editor} {\bibfnamefont {K.}~\bibnamefont {Hirakawa}}, \ and\ \bibinfo {editor} {\bibfnamefont {H.}~\bibnamefont {Yamaguchi}}}\ (\bibinfo  {publisher} {Springer Nature Singapore},\ \bibinfo {address} {Singapore},\ \bibinfo {year} {2022})\ pp.\ \bibinfo {pages} {257--276}\BibitemShut {NoStop}%
\bibitem [{\citenamefont {Cooper}\ and\ \citenamefont {Tripathi}(2008)}]{Cooper}%
  \BibitemOpen
  \bibfield  {author} {\bibinfo {author} {\bibfnamefont {N.~R.}\ \bibnamefont {Cooper}}\ and\ \bibinfo {author} {\bibfnamefont {V.}~\bibnamefont {Tripathi}},\ }\href {\doibase 10.1103/PhysRevB.77.245324} {\bibfield  {journal} {\bibinfo  {journal} {Phys. Rev. B}\ }\textbf {\bibinfo {volume} {77}},\ \bibinfo {pages} {245324} (\bibinfo {year} {2008})}\BibitemShut {NoStop}%
\bibitem [{\citenamefont {Tycko}\ \emph {et~al.}(1995)\citenamefont {Tycko}, \citenamefont {Barrett}, \citenamefont {Dabbagh}, \citenamefont {Pfeiffer},\ and\ \citenamefont {West}}]{Tycko1460}%
  \BibitemOpen
  \bibfield  {author} {\bibinfo {author} {\bibfnamefont {R.}~\bibnamefont {Tycko}}, \bibinfo {author} {\bibfnamefont {S.}~\bibnamefont {Barrett}}, \bibinfo {author} {\bibfnamefont {G.}~\bibnamefont {Dabbagh}}, \bibinfo {author} {\bibfnamefont {L.}~\bibnamefont {Pfeiffer}}, \ and\ \bibinfo {author} {\bibfnamefont {K.}~\bibnamefont {West}},\ }\href {\doibase 10.1126/science.7539550} {\bibfield  {journal} {\bibinfo  {journal} {Science}\ }\textbf {\bibinfo {volume} {268}},\ \bibinfo {pages} {1460} (\bibinfo {year} {1995})}\BibitemShut {NoStop}%
\bibitem [{\citenamefont {Smet}\ \emph {et~al.}(2002)\citenamefont {Smet}, \citenamefont {Deutschmann}, \citenamefont {Ertl}, \citenamefont {Wegscheider}, \citenamefont {Abstreiter},\ and\ \citenamefont {von Klitzing}}]{Smet2002}%
  \BibitemOpen
  \bibfield  {author} {\bibinfo {author} {\bibfnamefont {J.~H.}\ \bibnamefont {Smet}}, \bibinfo {author} {\bibfnamefont {R.~A.}\ \bibnamefont {Deutschmann}}, \bibinfo {author} {\bibfnamefont {F.}~\bibnamefont {Ertl}}, \bibinfo {author} {\bibfnamefont {W.}~\bibnamefont {Wegscheider}}, \bibinfo {author} {\bibfnamefont {G.}~\bibnamefont {Abstreiter}}, \ and\ \bibinfo {author} {\bibfnamefont {K.}~\bibnamefont {von Klitzing}},\ }\href {\doibase 10.1038/415281a} {\bibfield  {journal} {\bibinfo  {journal} {Nature}\ }\textbf {\bibinfo {volume} {415}},\ \bibinfo {pages} {281} (\bibinfo {year} {2002})}\BibitemShut {NoStop}%
\bibitem [{\citenamefont {Hashimoto}\ \emph {et~al.}(2002)\citenamefont {Hashimoto}, \citenamefont {Muraki}, \citenamefont {Saku},\ and\ \citenamefont {Hirayama}}]{Hashimoto}%
  \BibitemOpen
  \bibfield  {author} {\bibinfo {author} {\bibfnamefont {K.}~\bibnamefont {Hashimoto}}, \bibinfo {author} {\bibfnamefont {K.}~\bibnamefont {Muraki}}, \bibinfo {author} {\bibfnamefont {T.}~\bibnamefont {Saku}}, \ and\ \bibinfo {author} {\bibfnamefont {Y.}~\bibnamefont {Hirayama}},\ }\href {\doibase 10.1103/PhysRevLett.88.176601} {\bibfield  {journal} {\bibinfo  {journal} {Phys. Rev. Lett.}\ }\textbf {\bibinfo {volume} {88}},\ \bibinfo {pages} {176601} (\bibinfo {year} {2002})}\BibitemShut {NoStop}%
\bibitem [{\citenamefont {Kumada}\ \emph {et~al.}(2006)\citenamefont {Kumada}, \citenamefont {Muraki},\ and\ \citenamefont {Hirayama}}]{Kumada2006}%
  \BibitemOpen
  \bibfield  {author} {\bibinfo {author} {\bibfnamefont {N.}~\bibnamefont {Kumada}}, \bibinfo {author} {\bibfnamefont {K.}~\bibnamefont {Muraki}}, \ and\ \bibinfo {author} {\bibfnamefont {Y.}~\bibnamefont {Hirayama}},\ }\href {\doibase 10.1126/science.1127094} {\bibfield  {journal} {\bibinfo  {journal} {Science}\ }\textbf {\bibinfo {volume} {313}},\ \bibinfo {pages} {329} (\bibinfo {year} {2006})}\BibitemShut {NoStop}%
\bibitem [{\citenamefont {Guo}\ \emph {et~al.}(2010)\citenamefont {Guo}, \citenamefont {Hao}, \citenamefont {Tu}, \citenamefont {Zhao}, \citenamefont {Lin}, \citenamefont {Cao}, \citenamefont {Li}, \citenamefont {Zhou}, \citenamefont {Guo},\ and\ \citenamefont {Jiang}}]{Guo}%
  \BibitemOpen
  \bibfield  {author} {\bibinfo {author} {\bibfnamefont {G.~P.}\ \bibnamefont {Guo}}, \bibinfo {author} {\bibfnamefont {X.~J.}\ \bibnamefont {Hao}}, \bibinfo {author} {\bibfnamefont {T.}~\bibnamefont {Tu}}, \bibinfo {author} {\bibfnamefont {Y.~J.}\ \bibnamefont {Zhao}}, \bibinfo {author} {\bibfnamefont {Z.~R.}\ \bibnamefont {Lin}}, \bibinfo {author} {\bibfnamefont {G.}~\bibnamefont {Cao}}, \bibinfo {author} {\bibfnamefont {H.~O.}\ \bibnamefont {Li}}, \bibinfo {author} {\bibfnamefont {C.}~\bibnamefont {Zhou}}, \bibinfo {author} {\bibfnamefont {G.~C.}\ \bibnamefont {Guo}}, \ and\ \bibinfo {author} {\bibfnamefont {H.~W.}\ \bibnamefont {Jiang}},\ }\href {\doibase 10.1103/PhysRevB.81.041306} {\bibfield  {journal} {\bibinfo  {journal} {Phys. Rev. B}\ }\textbf {\bibinfo {volume} {81}},\ \bibinfo {pages} {041306} (\bibinfo {year} {2010})}\BibitemShut {NoStop}%
\bibitem [{\citenamefont {Rhone}\ \emph {et~al.}(2015)\citenamefont {Rhone}, \citenamefont {Tiemann},\ and\ \citenamefont {Muraki}}]{Rhone}%
  \BibitemOpen
  \bibfield  {author} {\bibinfo {author} {\bibfnamefont {T.~D.}\ \bibnamefont {Rhone}}, \bibinfo {author} {\bibfnamefont {L.}~\bibnamefont {Tiemann}}, \ and\ \bibinfo {author} {\bibfnamefont {K.}~\bibnamefont {Muraki}},\ }\href {\doibase 10.1103/PhysRevB.92.041301} {\bibfield  {journal} {\bibinfo  {journal} {Phys. Rev. B}\ }\textbf {\bibinfo {volume} {92}},\ \bibinfo {pages} {041301} (\bibinfo {year} {2015})}\BibitemShut {NoStop}%
\bibitem [{\citenamefont {Tong}\ \emph {et~al.}(2015)\citenamefont {Tong}, \citenamefont {Friess}, \citenamefont {Yong-Qing}, \citenamefont {Shi-Shen}, \citenamefont {Umansky}, \citenamefont {Klitzing},\ and\ \citenamefont {H.~Smet}}]{Guan_2015}%
  \BibitemOpen
  \bibfield  {author} {\bibinfo {author} {\bibfnamefont {G.}~\bibnamefont {Tong}}, \bibinfo {author} {\bibfnamefont {B.}~\bibnamefont {Friess}}, \bibinfo {author} {\bibfnamefont {L.}~\bibnamefont {Yong-Qing}}, \bibinfo {author} {\bibfnamefont {Y.}~\bibnamefont {Shi-Shen}}, \bibinfo {author} {\bibfnamefont {V.}~\bibnamefont {Umansky}}, \bibinfo {author} {\bibfnamefont {K.~v.}\ \bibnamefont {Klitzing}}, \ and\ \bibinfo {author} {\bibfnamefont {J.}~\bibnamefont {H.~Smet}},\ }\href {\doibase 10.1088/1674-1056/24/6/067302} {\bibfield  {journal} {\bibinfo  {journal} {Chinese Physics B}\ }\textbf {\bibinfo {volume} {24}},\ \bibinfo {pages} {067302} (\bibinfo {year} {2015})}\BibitemShut {NoStop}%
\bibitem [{\citenamefont {Kobayashi}\ \emph {et~al.}(2011)\citenamefont {Kobayashi}, \citenamefont {Kumada}, \citenamefont {Ota}, \citenamefont {Sasaki},\ and\ \citenamefont {Hirayama}}]{Kobayashi}%
  \BibitemOpen
  \bibfield  {author} {\bibinfo {author} {\bibfnamefont {T.}~\bibnamefont {Kobayashi}}, \bibinfo {author} {\bibfnamefont {N.}~\bibnamefont {Kumada}}, \bibinfo {author} {\bibfnamefont {T.}~\bibnamefont {Ota}}, \bibinfo {author} {\bibfnamefont {S.}~\bibnamefont {Sasaki}}, \ and\ \bibinfo {author} {\bibfnamefont {Y.}~\bibnamefont {Hirayama}},\ }\href {\doibase 10.1103/PhysRevLett.107.126807} {\bibfield  {journal} {\bibinfo  {journal} {Phys. Rev. Lett.}\ }\textbf {\bibinfo {volume} {107}},\ \bibinfo {pages} {126807} (\bibinfo {year} {2011})}\BibitemShut {NoStop}%
\bibitem [{\citenamefont {Kawamura}\ \emph {et~al.}(2013)\citenamefont {Kawamura}, \citenamefont {Gottwald}, \citenamefont {Ono}, \citenamefont {Machida},\ and\ \citenamefont {Kono}}]{Kawamura2013}%
  \BibitemOpen
  \bibfield  {author} {\bibinfo {author} {\bibfnamefont {M.}~\bibnamefont {Kawamura}}, \bibinfo {author} {\bibfnamefont {D.}~\bibnamefont {Gottwald}}, \bibinfo {author} {\bibfnamefont {K.}~\bibnamefont {Ono}}, \bibinfo {author} {\bibfnamefont {T.}~\bibnamefont {Machida}}, \ and\ \bibinfo {author} {\bibfnamefont {K.}~\bibnamefont {Kono}},\ }\href {\doibase 10.1103/PhysRevB.87.081303} {\bibfield  {journal} {\bibinfo  {journal} {Phys. Rev. B}\ }\textbf {\bibinfo {volume} {87}},\ \bibinfo {pages} {081303} (\bibinfo {year} {2013})}\BibitemShut {NoStop}%
\bibitem [{\citenamefont {Fauzi}\ \emph {et~al.}(2018)\citenamefont {Fauzi}, \citenamefont {Noorhidayati}, \citenamefont {Sahdan}, \citenamefont {Sato}, \citenamefont {Nagase},\ and\ \citenamefont {Hirayama}}]{Fauzi2018}%
  \BibitemOpen
  \bibfield  {author} {\bibinfo {author} {\bibfnamefont {M.~H.}\ \bibnamefont {Fauzi}}, \bibinfo {author} {\bibfnamefont {A.}~\bibnamefont {Noorhidayati}}, \bibinfo {author} {\bibfnamefont {M.~F.}\ \bibnamefont {Sahdan}}, \bibinfo {author} {\bibfnamefont {K.}~\bibnamefont {Sato}}, \bibinfo {author} {\bibfnamefont {K.}~\bibnamefont {Nagase}}, \ and\ \bibinfo {author} {\bibfnamefont {Y.}~\bibnamefont {Hirayama}},\ }\href {\doibase 10.1103/PhysRevB.97.201412} {\bibfield  {journal} {\bibinfo  {journal} {Phys. Rev. B}\ }\textbf {\bibinfo {volume} {97}},\ \bibinfo {pages} {201412} (\bibinfo {year} {2018})}\BibitemShut {NoStop}%
\bibitem [{\citenamefont {Shailos}\ \emph {et~al.}(2006)\citenamefont {Shailos}, \citenamefont {Bird}, \citenamefont {Lilly}, \citenamefont {Reno},\ and\ \citenamefont {Simmons}}]{Shailos}%
  \BibitemOpen
  \bibfield  {author} {\bibinfo {author} {\bibfnamefont {A.}~\bibnamefont {Shailos}}, \bibinfo {author} {\bibfnamefont {J.~P.}\ \bibnamefont {Bird}}, \bibinfo {author} {\bibfnamefont {M.~P.}\ \bibnamefont {Lilly}}, \bibinfo {author} {\bibfnamefont {J.~L.}\ \bibnamefont {Reno}}, \ and\ \bibinfo {author} {\bibfnamefont {J.~A.}\ \bibnamefont {Simmons}},\ }\href {\doibase 10.1088/0953-8984/18/12/009} {\bibfield  {journal} {\bibinfo  {journal} {Journal of Physics: Condensed Matter}\ }\textbf {\bibinfo {volume} {18}},\ \bibinfo {pages} {3277} (\bibinfo {year} {2006})}\BibitemShut {NoStop}%
\bibitem [{\citenamefont {Rössler}\ \emph {et~al.}(2011)\citenamefont {Rössler}, \citenamefont {Baer}, \citenamefont {de~Wiljes}, \citenamefont {Ardelt}, \citenamefont {Ihn}, \citenamefont {Ensslin}, \citenamefont {Reichl},\ and\ \citenamefont {Wegscheider}}]{Rossler}%
  \BibitemOpen
  \bibfield  {author} {\bibinfo {author} {\bibfnamefont {C.}~\bibnamefont {Rössler}}, \bibinfo {author} {\bibfnamefont {S.}~\bibnamefont {Baer}}, \bibinfo {author} {\bibfnamefont {E.}~\bibnamefont {de~Wiljes}}, \bibinfo {author} {\bibfnamefont {P.-L.}\ \bibnamefont {Ardelt}}, \bibinfo {author} {\bibfnamefont {T.}~\bibnamefont {Ihn}}, \bibinfo {author} {\bibfnamefont {K.}~\bibnamefont {Ensslin}}, \bibinfo {author} {\bibfnamefont {C.}~\bibnamefont {Reichl}}, \ and\ \bibinfo {author} {\bibfnamefont {W.}~\bibnamefont {Wegscheider}},\ }\href {\doibase 10.1088/1367-2630/13/11/113006} {\bibfield  {journal} {\bibinfo  {journal} {New Journal of Physics}\ }\textbf {\bibinfo {volume} {13}},\ \bibinfo {pages} {113006} (\bibinfo {year} {2011})}\BibitemShut {NoStop}%
\bibitem [{\citenamefont {Tsitsishvili}\ and\ \citenamefont {Ezawa}(2005)}]{Ezawa}%
  \BibitemOpen
  \bibfield  {author} {\bibinfo {author} {\bibfnamefont {G.}~\bibnamefont {Tsitsishvili}}\ and\ \bibinfo {author} {\bibfnamefont {Z.~F.}\ \bibnamefont {Ezawa}},\ }\href {\doibase 10.1103/PhysRevB.72.115306} {\bibfield  {journal} {\bibinfo  {journal} {Phys. Rev. B}\ }\textbf {\bibinfo {volume} {72}},\ \bibinfo {pages} {115306} (\bibinfo {year} {2005})}\BibitemShut {NoStop}%
\bibitem [{\citenamefont {Chida}\ \emph {et~al.}(2012)\citenamefont {Chida}, \citenamefont {Hashisaka}, \citenamefont {Yamauchi}, \citenamefont {Nakamura}, \citenamefont {Arakawa}, \citenamefont {Machida}, \citenamefont {Kobayashi},\ and\ \citenamefont {Ono}}]{Chida2012}%
  \BibitemOpen
  \bibfield  {author} {\bibinfo {author} {\bibfnamefont {K.}~\bibnamefont {Chida}}, \bibinfo {author} {\bibfnamefont {M.}~\bibnamefont {Hashisaka}}, \bibinfo {author} {\bibfnamefont {Y.}~\bibnamefont {Yamauchi}}, \bibinfo {author} {\bibfnamefont {S.}~\bibnamefont {Nakamura}}, \bibinfo {author} {\bibfnamefont {T.}~\bibnamefont {Arakawa}}, \bibinfo {author} {\bibfnamefont {T.}~\bibnamefont {Machida}}, \bibinfo {author} {\bibfnamefont {K.}~\bibnamefont {Kobayashi}}, \ and\ \bibinfo {author} {\bibfnamefont {T.}~\bibnamefont {Ono}},\ }\href {\doibase 10.1103/PhysRevB.85.041309} {\bibfield  {journal} {\bibinfo  {journal} {Phys. Rev. B}\ }\textbf {\bibinfo {volume} {85}},\ \bibinfo {pages} {041309} (\bibinfo {year} {2012})}\BibitemShut {NoStop}%
\bibitem [{\citenamefont {Kawamura}\ \emph {et~al.}(2015)\citenamefont {Kawamura}, \citenamefont {Ono}, \citenamefont {Stano}, \citenamefont {Kono},\ and\ \citenamefont {Aono}}]{Kawamura2015}%
  \BibitemOpen
  \bibfield  {author} {\bibinfo {author} {\bibfnamefont {M.}~\bibnamefont {Kawamura}}, \bibinfo {author} {\bibfnamefont {K.}~\bibnamefont {Ono}}, \bibinfo {author} {\bibfnamefont {P.}~\bibnamefont {Stano}}, \bibinfo {author} {\bibfnamefont {K.}~\bibnamefont {Kono}}, \ and\ \bibinfo {author} {\bibfnamefont {T.}~\bibnamefont {Aono}},\ }\href {\doibase 10.1103/PhysRevLett.115.036601} {\bibfield  {journal} {\bibinfo  {journal} {Phys. Rev. Lett.}\ }\textbf {\bibinfo {volume} {115}},\ \bibinfo {pages} {036601} (\bibinfo {year} {2015})}\BibitemShut {NoStop}%
\end{thebibliography}

\begin{thebibliography}{99}

\bibitem{ARLong} A.R. Long and M. Pioro-Ladrière and J.H. Davies and A.S. Sachrajda and Louis Gaudreau and P. Zawadzki and J. Lapointe and J. Gupta and Z. Wasilewski and S.A. Studenikin, Physica E: Low-dimensional Systems and Nanostructures \textbf{34}, 553 (2006).





\end{thebibliography}
\end{document}